February 2023 • Dr Anna-Katharina Meßmer & Dr Martin Degeling

# Auditing Recommender Systems

## Putting the DSA into practice with a risk-scenario-based approach

# Executive Summary

Today's online platforms rely heavily on recommendation systems to serve content to their users; social media is a prime example. In turn, recommendation systems largely depend on artificial intelligence algorithms to decide who gets to see what. While the content social media platforms deliver is as varied as the users who engage with them, it has been shown that platforms can contribute to serious harm to individuals, groups and societies. Studies have suggested that these negative impacts range from worsening an individual's mental health to driving society-wide polarisation capable of putting democracies at risk.

To better safeguard people from these harms, the European Union's Digital Services Act (DSA) requires platforms, especially those with large numbers of users, to make their algorithmic systems more transparent and follow due diligence obligations. These requirements constitute an important legislative step towards mitigating the systemic risks posed by online platforms. However, the DSA lacks concrete guidelines to operationalise a viable audit process that would allow auditors to hold these platforms accountable. This void could foster the spread of 'audit-washing', that is, platforms exploiting audits to legitimise their practices and neglect responsibility.

To fill this gap, we propose a **risk-scenario-based audit process**. We explain in detail what audits and assessments of recommender systems according to the DSA should look like. Our approach also considers the evolving nature of platforms and emphasises the observability of their recommender systems' components. The resulting audit facilitates internal (among audits of the same system at different moments in time) and external comparability (among audits of different platforms) while also affording the evaluation of mitigation measures implemented by the platforms themselves.

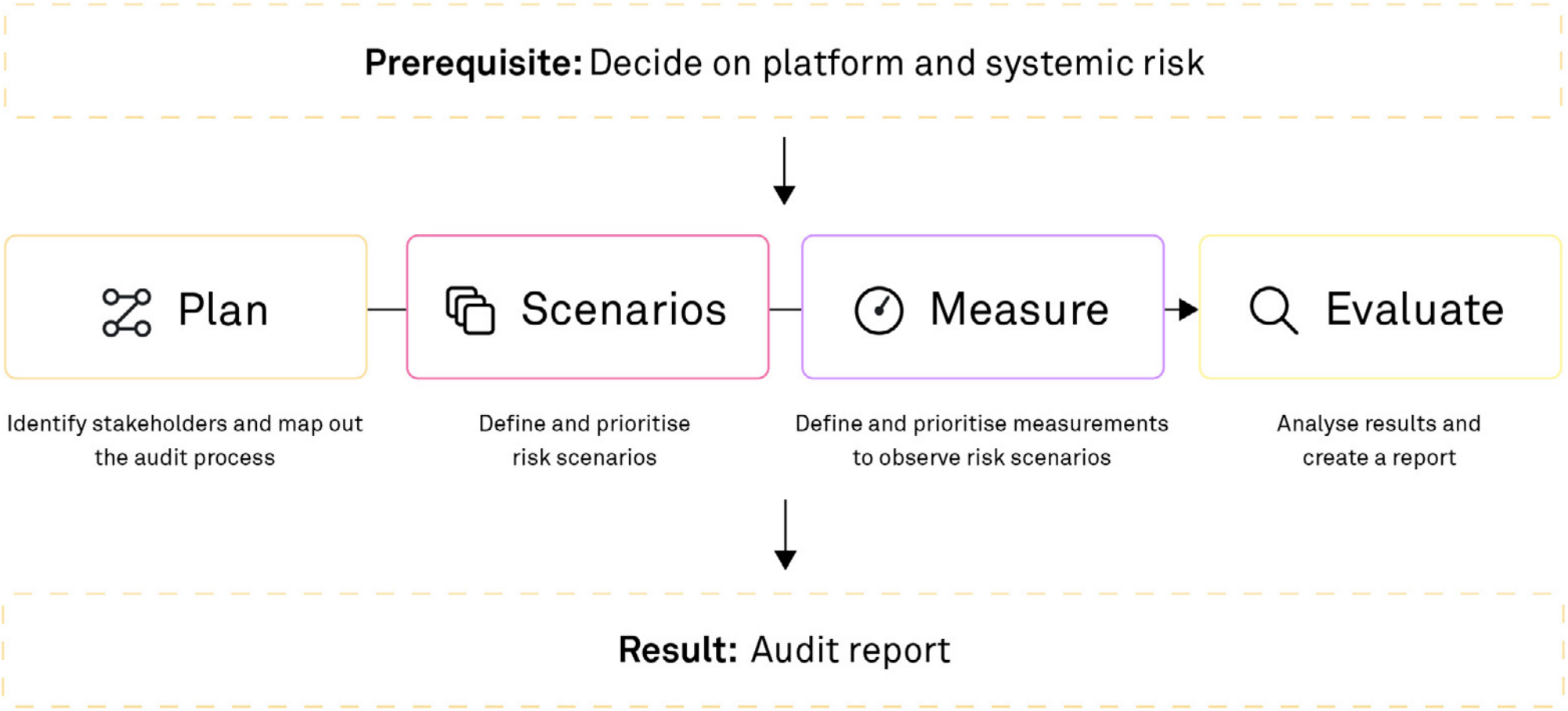

**The risk-scenario-based audit process consists of four consecutive steps:**

**1. Plan:** The first step aims to obtain a deep understanding of the platform under scrutiny and use this information to determine the profiles of stakeholders who should be involved in the audit process. Depending on the experience and expertise needed, this could include platform developers, researchers, legal experts and representatives of parties affected.

**2. Define Scenarios:** Step two focuses on the definition and prioritisation of the scenarios. A scenario is a description of specific issues related to a systemic risk. The scenario breaks down abstract risks like harms to 'mental well-being' into concrete testable hypotheses by defining the affected party and its characteristics (e.g., a young adult in a personal crisis), the harm (e.g., being excessively exposed to content describing or showing self-harm), the involved elements of the platform (e.g., the recommender system of a social media timeline) and the further impact (e.g., increase of mental health crisis of young adults). A systemic risk may often involve several scenarios; therefore, prioritising and selecting is necessary.

**3. Measure:** The next step involves developing measurements to understand the scenario in the context of the platform. There are many different approaches, both in terms of the test method and elements of the platform under scrutiny. Approaches can range from automated measurements that look at the actual code or the spread of specific content to user perspectives through surveys. An auditor needs to develop multiple measurements and then prioritise them to find the best one(s) to test a specific scenario.

**4. Evaluate:** The final step of the process is to analyse the results and write an audit report. The report should enable reproducibility and recommend mitigation measures.

Without such a structured, multi-stakeholder auditing process, the DSA's potentially innovative rules on audits and risk assessments will be rendered useless. The European Commission as well as national regulatory bodies can benefit from studying and adapting the proposed steps for their DSA enforcement efforts: It would help regulators build up expertise and networks on platform risks, which is necessary when dealing with professional, for-profit auditors and the audited companies themselves. Moreover, the guidelines we propose also provide any independent party with a roadmap to carry out an audit, thus contributing to a more integral societal oversight of the recommender systems of these platforms.

# Table of Contents

# 1. Introduction

What would the world look like today without social media platforms? Many people can hardly imagine a life without them because these platforms have become an essential part of people's daily lives. For instance, news consumption is quickly moving to the social platform arena, and European users are no exception. According to the Reuters Digital News Report, online and social media have overtaken television as a news source in Germany for the first time in 2022.[1]

Social media platforms provide users with connectivity and community, information and structures for social impact and political change, but these digital environments also carry the risk of adverse effects. Studies have shown that social media can negatively affect individuals, groups and societies.[2] It has been suggested that social media may be not only responsible for news fatigue, news avoidance or reduced well-being at the individual level, but also pose significant threats to the integrity of democracies.[3] Moreover, the opaqueness of the inner workings of these technologies and the algorithms they use makes it challenging to achieve a more comprehensive understanding of their potential societal effects.

Despite the far-reaching risks that social platforms carry, policymakers have long failed to tackle the challenges posed by these technologies, even seeming reluctant to regulate them. However, this has changed—at least within the European Union (EU). The 'Digital Services Act'[4](DSA) is a critical step in regulating social media platforms. After years of preparations and two years of negotiations within the EU and the public, the DSA went into force in October 2022, and its rules will apply from 2024 onwards. The DSA regulates the obligations of intermediaries (i.e., social media or video-sharing platforms, search engines, hosting providers and marketplaces) to protect

1 Nic Newman et al., 'Reuters Institute Digital News Report 2022' (Reuters Institute for the Study of Journalism, 2022), 12, https://reutersinstitute.politics.ox.ac.uk/sites/default/files/2022-06/Digital_News-Report_2022.pdf.

2 Jonathan Haidt and Chris Bail, 'Social Media and Political Dysfunction: A Review', *New York University*, 2022, https://docs.google.com/document/d/1vVAtMCQnz8WVxtSNQev_e1cGmY9rnY96ecYuAj6C548/; Jonathan Haidt and Jean Twenge, 'Social Media Use and Mental Health: A Review', *New York University*, 2021, https://docs.google.com/document/d/1w-HOfseF2wF9YIpXwUUtP65-olnkPyWcgF5BiAtBEy0/

3 WSJ, 'Inside TikTok's Algorithm: A WSJ Video Investigation', *Wall Street Journal*, 21 July 2021, https://www.wsj.com/articles/tiktok-algorithm-video-investigation-11626877477; Jesse McCrosky and Brandi Geurkink, 'YouTube Regrets: A Crowdsourced Investigation into YouTube's Recommendation Algorithm' (Brussels: Mozilla Foundation, 7 July 2021), https://assets.mofoprod.net/network/documents/Mozilla_YouTube_Regrets_Report.pdf; Johanne Kübler et al., 'The 2021 German Federal Election on Social Media: An Analysis of Systemic Electoral Risks Created by Twitter and Facebook Based on the Proposed EU Digital Services Act' (Sustainable Computing Lab, August 2021), https://www.sustainablecomputing.eu/wp-content/uploads/2021/10/DE_Elections_Report_Final_17.pdf.

4 European Union, 'Regulation (EU) 2022/2065 of the European Parliament and of the Council on a Single Market For Digital Services (Digital Services Act) and Amending Directive 2000/31/EC' (2022), https://eur-lex.europa.eu/legal-content/EN/TXT/?uri=CELEX:32022R2065.

users and their fundamental rights online by establishing a transparency and accountability framework.[5]

In its attempt to limit negative effects, one focus of the DSA is requirements for the providers of digital platforms, especially due diligence requirements for 'very large online platforms' (VLOPs; usually including very large online search engines)[6], to reduce the harms that their algorithm-based systems may directly or indirectly inflict on individuals, groups and societies. VLOPs should ensure this with increased transparency, reporting and audits before launching new products and keep doing so regularly during their existence.

Now that the final version of the DSA is in force, the upcoming months and years present the challenges posed by its implementation and enforcement. To evaluate whether and where intermediaries are in violation of their provisions, the DSA requires them to carry out a set of different audits and assessments. However, guidelines on how these audits must be conducted are missing and need to be defined over the upcoming months.[7] There are different implementing and delegated acts, most importantly on independent audits by external contractors, that regulators are developing. If civil society does not support the development of standards, definitions and concrete procedures, the platforms will be left to their own (self-serving) assessments. This is why, in the current paper, we take a closer look at this central aspect of the DSA: audits and assessments of the potential harms and risks of AI-based recommender systems of social media platforms. In the following sections, we put forward a suitable process that is useful for all potential auditors.

We propose an **audit process** based on **risk scenarios**. The process is organised around scenarios that describe the specific negative effects of a recommender system in a concrete and systematic form. As part of this process, various platform elements, issues, stakeholders and audit types need to be considered regarding this specific risk scenario. We map the process steps to requirements established in the DSA and provide auditors with guidelines for making a well-founded decision on a suitable audit or assessment method. The process enables comparability between different approaches, audits and assessments by systematising the scenarios and several of the components involved. This approach is aimed

5 European Commission, 'Questions and Answers: Digital Services Act', Text, European Commission - European Commission, 14 November 2022, https://ec.europa.eu/commission/presscorner/detail/en/QANDA_20_2348.
6 Very large online platforms are defined as having more than 45 million (a number equivalent to 10% of the European Union population) active recipients, which is calculated as an average over a period of six months (see Recital 76, DSA).
7 eCommerce Experts Group, '25th ECommerce Expert Group Meeting', https://ec.europa.eu/transparency/expert-groups-register/screen/meetings/consult?lang=en&meetingId=45535.

at the observability of platforms, which means drawing attention to the process dimension and practice of observing platforms[8] that are constantly evolving. Therefore, this approach is also useful for assessing whether any implemented mitigation measures have the desired effect.

Before we introduce the process step by step (Section 5) and give an outlook (Section 10), we first elaborate on the following questions: What are recommender systems, and how do they work? What are the systemic risks the DSA mentions, and how can they be operationalised (Section 3)? We also look at the definitions of audits and assessments within the DSA and the literature (Section 4).

## 2. What are recommender systems?

Social media platforms are bustling with activity. Any given minute, 500 hours of video are uploaded to YouTube; 66,000 pictures are shared on Instagram; 1,700,000 pieces of content are shared on Facebook; 347,200 tweets are shared on Twitter;[9] and 167,000,000 million videos are watched on TikTok.[10] To make sense of these vast amounts of information—and to optimise for user engagement as part of the prevailing business models—the recommender systems of platforms are constantly sorting, filtering and selecting the content that will be presented to users. Recommender systems are one of the main factors ultimately determining what specific content users end up spending their time and interacting with. Therefore, these systems constitute one of the most crucial components of the platforms that the DSA regulates. The Digital Services Act defines a recommender system as follows:

> *'"recommender system" means a fully or partially automated system used by an online platform to suggest in its online interface specific information to recipients of the service or prioritise that information, including as a result of a search initiated by the recipient of the service or otherwise determining the relative order or prominence of information displayed.' (Article 3(s), DSA)*

The DSA focuses on the audits and risk assessments of very large social media and video-sharing platforms like Facebook, Instagram, TikTok, Twitter and YouTube, including the assessment of their recommender systems. However,

8 Bernhard Rieder and Jeanette Hofmann, 'Towards Platform Observability', *Internet Policy Review* 9, no. 4 (18 December 2020), https://policyreview.info/articles/analysis/towards-platform-observability.
9 Domo, 'Data Never Sleeps 10.0', 2022, https://www.domo.com/data-never-sleeps.
10 Domo, 'Data Never Sleeps 9.0', 2021, https://www.domo.com/data-never-sleeps.

size is not the only factor that determines a recommender system's impact. It is also a question of how content can be shared, how users can interact with it and how it is processed by the recommender systems. With this in mind, our approach has been developed to afford its applicability to any relevant recommender system, regardless of its size. In the following section, we elaborate on two important aspects of how recommender systems work: affordances and choice architecture (Section 2.1) and the different parts and stages of a recommender system (Section 2.2).

## 2.1. Choice architecture and affordances

To achieve its goals, every platform is developed with a particular structure, set of features, design philosophy, language, terms of service and so forth in mind.[11] From the user experience perspective, the set of elements that enables or forbids certain actions, nudges users to specific behaviour and makes a desired behaviour and certain choices more likely is referred to as a system's 'choice architecture'[12] or 'affordances'.[13] The affordances of a platform affect the content itself (e.g., by setting boundaries for the length of a video), the user's behaviour (e.g., by offering different settings for the structure of a timeline) and the interaction between content and user (e.g., by not offering a certain kind of feedback). Therefore, a platform's recommender system is highly dependent on or intertwined with its choice architecture.

For example, TikTok allowed only very short videos for a long time. Although the limit was recently extended, many videos are still short because this affordance has impacted how creators design their content. Moreover, many of TikTok's current features evolved from an app where users could reuse and remix trending music over their own videos. Competitors such as Instagram have picked up this unique feature of TikTok's choice architecture.

Another example illustrates the impact that changes to architecture can have. In 2016, Facebook added emoji 'reactions' to the like button, including

11 For the importance of platform design and user interface see: Julian Jaursch, 'Strengthening EU Proposals on Deceptive Platform Design: Ideas on How to Improve the Draft Digital Services Act' (Berlin: Stiftung Neue Verantwortung, 15 March 2022), https://www.stiftung-nv.de/sites/default/files/snv_strengthening_eu_proposals_on_deceptive_platform_design.pdf; Sebastian Rieger and Caroline Sinders, 'Dark Patterns: Regulating Digital Design' (Berlin: Stiftung Neue Verantwortung, 13 May 2020), https://www.stiftung-nv.de/sites/default/files/dark.patterns.english.pdf.

12 Stephan Lewandowsky et al., 'Technology and Democracy: Understanding the Influence of Online Technologies on Political Behaviour and Decision-Making' (Luxembourg: Publications Office of the European Union, 26 October 2020), 20, http://op.europa.eu/en/publication-detail/-/publication/49b629ee-1805-11eb-b57e-01aa75ed71a1/language-en.

13 Sandra K. Evans et al., 'Explicating Affordances: A Conceptual Framework for Understanding Affordances in Communication Research', *Journal of Computer-Mediated Communication* 22, no. 1 (2017): 35–52, https://doi.org/10.1111/jcc4.12180.

an angry emoji. From then on, the user interface offered users more nuanced reactions to content, which, in turn, made it easier to algorithmically differentiate between the users' engagement. Before this change, user feedback by clicking the like button could only be measured as a binary value: it was either one, meaning the user clicked the like button and had shown a response to the post, or zero, indicating that either the user had no response or did not 'like' the post. The diversification of reactions meant that the recommender could use more data points, but reactions also seemed to be a better indicator of attention and interaction. Therefore, Facebook started to weigh emojis five times higher than the like button in the mathematical calculation of the recommendation algorithms.[14] Although this was a change to only one of many user engagement variables and their numerous calculations, this development seemed to have an enormous impact. It was criticised for spreading more misinformation and violent content across the platform and was eventually backtracked—with the result being that users received less problematic content.

It is important to emphasise that platform affordances make certain user behaviour and choices more probable but do not determine it entirely. Therefore, the analysis of affordances must always be accompanied by an analysis of actual use. As Mark Eisenegger, professor of the public sphere and society, put it, 'Platform logics are constituted [...] in the interdependence of platform affordances (1), the actual use (2) of these affordances, and the algorithms (3) that mediate between the two dimensions'.[15] For this reason, platform affordances and elements of the choice architecture, like the user interface shaping the user experience, need to be considered as part of the audit process of a recommendation algorithm. That takes us to the second aspect: 'the' algorithm.

## 2.2. Why it is not 'the' algorithm, but many

There is a common misconception about how recommender systems work: 'the' algorithm. However, platforms do not rely on a single, all-encompassing algorithm that handles every recommendation decision. Platform choice architectures are usually divided into different technical products and

14 Jeremy B. Merrill and Will Oremus, 'Five Points for Anger, One for a "like": How Facebook's Formula Fostered Rage and Misinformation', *Washington Post*, 26 October 2021, https://www.washingtonpost.com/technology/2021/10/26/facebook-angry-emoji-algorithm/.

15 Translated from german: Mark Eisenegger, 'Dritter, digitaler Strukturwandel der Öffentlichkeit als Folge der Plattformisierung', in Digitaler Strukturwandel der Öffentlichkeit: *Historische Verortung, Modelle und Konsequenzen*, ed. Mark Eisenegger et al., Mediensymposium (Wiesbaden: Springer Fachmedien, 2021), 26, https://doi.org/10.1007/978-3-658-32133-8_2.

functionalities in which several algorithms work together. Reducing a platform's choice architecture to 'the' algorithm obscures the nuances and intricacies of these systems.

For example, when a video is uploaded to TikTok, it goes through several, mostly automated, steps before being published and before being recommended to the user's 'For You' feed. First, a system will scan it for known issues, such as containing offensive or illegal material. If it passes this threshold, it will then be distributed to a small number of users to obtain some preliminary information about it and overcome what is known as the cold start problem: the inability to filter content for which not enough information is yet available. Once more data are available, a custom recommender can decide whether the video is suitable for a specific user. If a particular video 'goes viral', it will then be distributed to even more users and enter different pools of heavily pushed—and hence viewed—videos.[16]

Differentiation into technical products can be quite sophisticated. We use YouTube as an example. Even if you do not have any technical knowledge, you may have noticed at least six features that provide some form of content recommendation: the homepage, its search interface, the autoplay function at the end of a video, the section with recommendations based on the currently watched video and the newly launched 'shorts' section, all of which are intertwined with an advertisement system. All of these features are separate technical products with their own development team and objectives. Moreover, the recommendation systems of each of these products make use of several interlocking algorithms that often provide feedback to each other.[17]

When we discuss assessing the risks of recommender systems, we must not only decide which platform to analyse, but also which specific product on the platform should be evaluated. Even within the recommendation system of one product, several algorithms are interlocked and built on each other.[18]

Following Stray et al.,[19] the generalised process of a recommendation engine looks like Figure 1.

16 Catherine Wang, 'Why TikTok Made Its User so Obsessive? The AI Algorithm That Got You Hooked. | by Catherine Wang | Towards Data Science', *Towards Datascience* (blog), 7 January 2020, https://towardsdatascience.com/why-tiktok-made-its-user-so-obsessive-the-ai-algorithm-that-got-you-hooked-7895bb1ab423.

17 Anna Semenova, 'Upcoming Paper' (Berlin: Stiftung Neue Verantwortung, 2023).

18 Zhe Zhao et al., 'Recommending What Video to Watch next: A Multitask Ranking System', in *Proceedings of the 13th ACM Conference on Recommender Systems* (RecSys '19: Thirteenth ACM Conference on Recommender Systems, Copenhagen Denmark: ACM, 2019), 43–51, https://doi.org/10.1145/3298689.3346997.

19 Jonathan Stray et al., 'Building Human Values into Recommender Systems: An Interdisciplinary Synthesis' (arXiv, 20 July 2022), https://doi.org/10.48550/arXiv.2207.10192

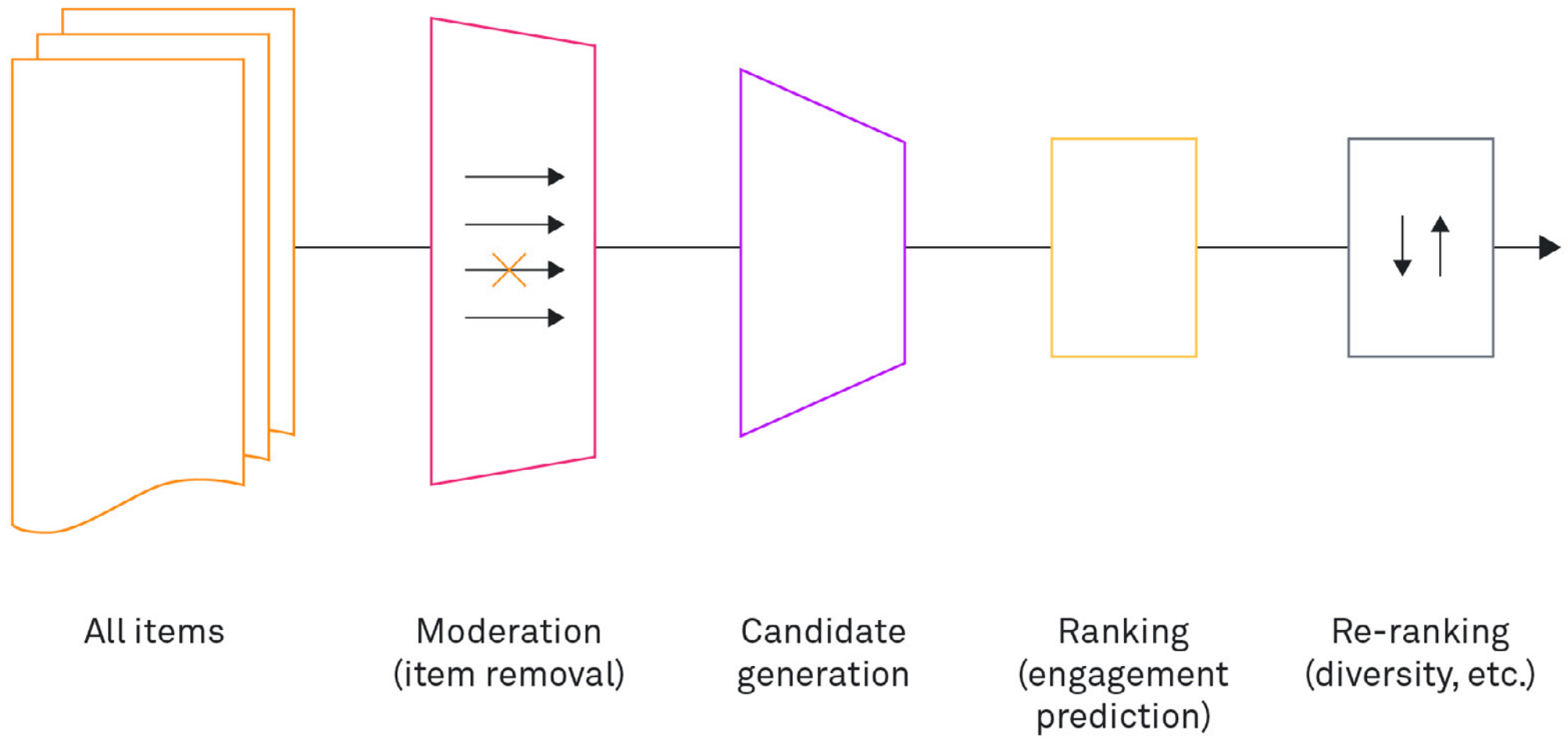

Figure 1:
An overview of the different steps in a recommendation system (taken from Stray et al., 2022)

As we can see in Figure 1[20], recommender systems and content moderation are highly intertwined. The first step of the recommender system is 'moderation', which means content may already be blocked by automated moderation at this stage—at least if it violates any aspect of the terms of service. The terms of service of a platform, as well as laws and regulations, influence what can and cannot be uploaded or what content is labelled (e.g., as disinformation and/or as so-called borderline content, leading to demotion in the following stages).

The next two steps after 'moderation' are 'candidate generation' and 'ranking'. YouTube developers describe these steps as follows: 'At the candidate generation stage, we retrieve a few hundred [video] candidates from a huge corpus [of videos]. Our ranking system provides a score for each [video] and generates the final ranked list'.[21] They specify the following:

> *'Our video recommendation system uses multiple candidate generation algorithms, each of which captures one aspect of similarity between query video and candidate video. For example, one algorithm generates candidates by matching topics of query video. Another algorithm retrieves candidate videos based on how often the video has been watched together with the query video. We construct a sequence model [...] for generating personalized candidate given user history. We also use techniques [...] to generate context-aware high recall relevant candidates.'*[22]

20 Stray et al., 14.
21 Zhao et al., 'Recommending What Video to Watch Next', 46.
22 Zhao et al., 46.

Here, it becomes clear how different algorithms interlock at different levels of the recommender system—for example, to create a list of candidate videos. Two processes run in parallel: first, the clustering of content—for example, according to specific topics—and second, the clustering of users—for example, according to specific interests. When we talk about algorithms as a black box, it is not so much the general mode of operation that is not transparent, but rather, it is the concrete clustering. What is clear, though, is which variables can go into the calculations: explicit feedback of (dis-) satisfaction, such as likes or blocking; positive and negative implicit signals of interest, such as the watch time of a video or skipping; other user data such as age, gender or the history of the account; usage statistics such as the software and hardware or time spent online; and the friendship connections or followings (the social graph). However, the importance of the latter is currently decreasing in social media. TikTok marks the transition from the social graph, as relevant on Twitter, Facebook and Mastodon, to the interest graph, which has also been adopted by Instagram and YouTube, at least in certain sections. In an influential article, Michal Mignano, an investor and former Spotify employee, marked the 'The End of Social Media and the Rise of Recommendation Media'.[23] For recommendations based on the interest graph, subscriptions, followings, 'friendships' or interactions with other users play a less critical role. Instead, the focus is on interests, which rely heavily on interactions with the content itself, especially watch time.[24] For example, the TikTok predictive model ties together user-related information, including which videos have been viewed recently, video-related information, such as the video duration, and video creator-related information, such as the historical statistics of how many videos the creator has uploaded and how often they were viewed and liked. This means that, in the final step of the recommendation system, probabilities and predictions are made based on historical data on past usage and connections. Different platforms also optimise their recommendation system for different metrics set by the companies (and their internal ranking teams), which might include watch time, time spent on the apps, likes and interaction or personalisation.

This shift from social to interest graphs makes recommendation engines even more influential because it limits the options for users to intervene directly, for example, by unfollowing certain accounts or choosing a chronologically sorted timeline.

23 Michael Mignano, 'The End of Social Media and the Rise of Recommendation Media', *Medium* (blog), 8 August 2022, https://mignano.medium.com/the-end-of-social-media-a88ffed21f86.

24 Ben Smith, 'How TikTok Reads Your Mind', *The New York Times*, 5 December 2021, https://www.nytimes.com/2021/12/05/business/media/tiktok-algorithm.html.

# 3. What are risks?

As shown above, a platform is never 'neutral', and there is no such thing as a 'neutral' social media timeline, even if it is sorted chronologically. As scientific and journalistic research shows, recommender systems pose several risks to individuals, groups and society. For example, the first stage of a recommender system (i.e., automated content moderation) already has its flaws, often failing to identify offensive or illegal material or overblocking content. This is why the Digital Services Act has a strong focus on the 'systemic risks' of platforms.

When evaluating recommendation engines, we think it is best to start with the following risks, as defined by the DSA, which lists four categories: (1) the dissemination of illegal content, (2) the impact on fundamental human rights, (3) negative effects on democratic processes and public security and (4) negative consequences on public health, minors, mental well-being or gender-based violence.

**The DSA mentions four categories of 'systemic risks':**

**1.** 'risks associated with the dissemination of illegal content, such as the dissemination of child sexual abuse material or illegal hate speech or other types of misuse of their services for criminal offenses, and the conduct of illegal activities'. (Recital 80)

**2.** 'the actual or foreseeable impact of the service on the exercise of fundamental rights, as protected by the Charter, including but not limited to human dignity, freedom of expression and of information, including media freedom and pluralism, the right to private life, data protection, the right to non-discrimination, the rights of the child and consumer protection.' (Recital 81)

**3.** 'the actual or foreseeable negative effects on democratic processes, civic discourse and electoral processes, as well as public security.' (Recital 82)

**4.** 'concerns relating to the design, functioning or use, including through manipulation, of very large online platforms and of very large online search engines with an actual or foreseeable negative effect on the protection of public health, minors and serious negative consequences to a person's physical and mental wellbeing, or on gender-based violence.' (Recital 83)

It is a challenge that this list names an enormous range of risks and impacts that vary in how well they are defined or can be defined at all. In addition, the risks mentioned also operate at very different levels. Some of these risks can be defined quite easily based on existing legislation, for example, when it comes to 'illegal activities', such as the sale of products or services prohibited by Union or national law'. Therefore, some risks, such as the distribution of child sexual abuse material, can be checked at the platform level by utilising databases provided by entities such as the National Center for Missing and Exploited Children (NCMEC), a U.S. child protection organisation.[25]

Other risks, however, are completely undefined, quite abstract and vague or different and sometimes contradictory definitions exist in different disciplines. This is especially true for all risks that are described as having an impact on individuals (e.g., psychological well-being), groups of individuals (nondiscrimination) or societies (civil society discourse and election processes). These risks depend on a complex socio-technical interplay between platform design, different algorithms, users and society. As social use practices change, algorithmic decision models change as well.[26] Therefore, the level at which these risks can be observed and evaluated varies widely, too. What is needed here is, first, much more concrete descriptions of the risk scenarios and, second, accompanying sociological and/or psychological research to capture the effects that take place beyond the platforms or in the interaction between platforms and users.

One very good example is the debate around 'filter bubbles',[27] 'echo chambers'[28] and polarisation on social media platforms. Whether these exist at all and to what extent social media causes them is still in discussion. Although these three risks are thought of as highly intertwined, they function and are observable on different levels.

Whereas the 'filter bubble' theory claims that algorithms push users into consuming an increasing number of homogeneous content and, therefore, making people become more polarised (technical level), the 'echo chamber' theory suggests that it is rather the human tendency to connect with like-minded people that causes polarisation (psychological level). To carry out

25 In the brevity of this example, we do not go into the details of the process, the technical drawbacks of the approach used by these databases or the critique of NCMEC as an organisation (Sebastian Meineck, 'NCMEC Figures Explained: How the Spectre of Millionfold Abuse Haunts European Policy-Makers', *netzpolitik.org* (blog), 20 July 2022, https://netzpolitik.org/2022/ncmec-figures-explained-how-the-spectre-of-millionfold-abuse-haunts-european-policy-makers/; Neal Krawetz, 'One Bad Apple', *The Hacker Factor Blog* (blog), 8 August 2021, https://www.hackerfactor.com/blog/index.php?/archives/929-One-Bad-Apple.html. for technical details).
26 Rieder and Hofmann, 'Towards Platform Observability'.
27 Eli Pariser, *The Filter Bubble: What the Internet Is Hiding from You* (New York: Penguin Press, 2011).
28 Cass R. Sunstein, *#Republic: Divided Democracy in the Age of Social Media*, Third printing, and first paperback printing (Princeton Oxford: Princeton University Press, 2018).

research on 'echo chambers', this individual preference for like-minded positions needs to be considered. Conversely, if the research interest is 'filter bubbles', it is necessary to exclude this individual factor to attribute impacts to the algorithm itself.

However, regardless of the perspective, current research has suggested that the existence of both phenomena was highly overestimated.[29] This raises the question of whether both issues were predicated on the wrong assumptions. If we revisit the question of what might cause polarisation in general, it is also possible that it is *exposure* to opposite views that could shape polarisation, not the *absence* of it.[30] Here, the challenge in distinguishing societal risks from their roots in social media platforms becomes even more apparent. First, polarisation is not a technical but a social and only sociologically or psychologically approachable risk, and second, it cannot be attributed solely to platform-related reasons. Therefore, we must understand systemic risks more as a socio-technical interplay.

This is why, when assessing these and other systemic risks, they must be studied in relation to the specifics of each platform and its products, as described above (e.g., the user interface or the different stages of a recommender system). This is a basic condition mandated by the DSA, at least for internal risk assessments of VLOPs (see Article 34). Besides this, it will be necessary for any risk assessment to differentiate between those risks that can be observed on a platform, for example, the *spread of content* about depression and suicide[31] or promoting eating disorders[32], and *the impact* these risks have on certain people, groups or society. Proving the spread of specific content does not necessarily have a causal relation to developments outside of the platform, for example, the mental well-being of individuals.[33]

29 Birgit Stark, Melanie Magin, and Pascal Jürgens, 'Maßlos überschätzt. Ein Überblick über theoretische Annahmen und empirische Befunde zu Filterblasen und Echokammern', in *Digitaler Strukturwandel der Öffentlichkeit: Historische Verortung, Modelle und Konsequenzen*, ed. Mark Eisenegger et al., Mediensymposium (Wiesbaden: Springer Fachmedien, 2021), 303–21, https://doi.org/10.1007/978-3-658-32133-8_17.

30 Petter Törnberg, 'How Digital Media Drive Affective Polarization through Partisan Sorting', *Proceedings of the National Academy of Sciences* 119, no. 42 (18 October 2022), https://doi.org/10.1073/pnas.2207159119; Christopher A. Bail et al., 'Exposure to Opposing Views on Social Media Can Increase Political Polarization', Proceedings of the National Academy of Sciences 115, no. 37 (11 September 2018): 9216–21, https://doi.org/10.1073/pnas.1804840115.

31 WSJ, 'Inside TikTok's Algorithm'

32 Angus Crawford, 'Instagram Eating Disorder Content 'out of Control'', *BBC News*, 20 March 2019, https://www.bbc.com/news/uk-47637377.

33 Sometimes, even the content itself is hard to define, for example, when it comes to disinformation. Naomi Appelman et al., 'Truth, Intention and Harm: Conceptual Challenges for Disinformation-Targeted Governance', 2022, https://policyreview.info/articles/news/truth-intention-and-harm-conceptual-challenges-disinformation-targeted-governance/1668; Ronan Ó Fathaigh, Natali Helberger, and Naomi Appelman, 'The Perils of Legally Defining Disinformation', *Internet Policy Review* 10, no. 4 (4 November 2021), https://policyreview.info/articles/analysis/perils-legally-defining-disinformation.

In this section, we have described the intricacies of the risks that recommendation systems pose to individuals, groups and societies. One thing is clear already: these risks cannot be assessed using a reductionist checklist approach. This paradigm falls short because the questions posed by these risks are significantly more complex and challenging than what can be captured by a list of yes/no checkboxes. This brings us to the next question: What are audits and assessments?

# 4. What are audits and assessments?

Audits and impact assessments are widely used in many fields as a method to evaluate the potential impacts of innovations and determine their compliance with law. Be it protecting sensitive user data or preserving the environment or safeguarding human rights, these assessment frameworks enable the systematic evaluation of a given system, intervention or process. Because authorities often issue guidance of a mostly general nature, other actors—such as local authorities—and stakeholders must operationalise and complement this oversight.

For example, the EU's General Data Protection Regulation (GDPR) requires anyone who collects sensitive personal information, such as biometric data, or who uses specific types of processing, such as scoring, to conduct a data protection impact assessment (DPIA). As a result, the data protection authorities have published lists of processes that require DPIAs. In turn, researchers, standardisation bodies and civil society organisations have published guidelines for data protection officers in organisations that describe protection goals and safeguards. First, the 'Article 29 Working Group' within the European Commission published general guidelines elaborating on the legal descriptions within the GDPR.[34] Based on these, data protection authorities at the national level, like the French 'Commission Nationale de l'Informatique et des Libertes' or the conference of the data protection authorities in Germany, have released more in-depth descriptions of the process.[35] Civil society organisations like the 'Forum Privatheit' in Germany then proposed processes for DPIAs that describe an approach in even more detail.[36]

34 Article 29 Working Party, 'Guidelines on Data Protection Impact Assessment (DPIA) and Determining Whether Processing is 'Likely to Result in a High Risk' for the Purposes of Regulation 2016/679' (European Commission, 4 April 2017), https://ec.europa.eu/newsroom/just/document.cfm?doc_id=47711.

35 CNIL, 'Guidelines on DPIA', 18 October 2017, https://www.cnil.fr/en/guidelines-dpia.

36 Michael Friedewald et al., 'Datenschutz-Folgenabschätzung: Ein Werkzeug Für Einen Besseren Datenschutz' (Karlsruhe: Forum Privatheit und selbstbestimmtes Leben in der digitalen Welt, 29 November 2017), https://www.forum-privatheit.de/wp-content/uploads/Forum-Privatheit-WP-DSFA-3-Auflage-2017-11-29.pdf.

Similarly, in the field of human rights, standardised tools, such as human rights impact assessments (HRIA), offer a methodology to carry out evaluations. Based on the United Nations Guiding Principles on Business and Human Rights, different civil society organisations have developed toolboxes and processes that make the implementation of HRIAs easier. A notable example is the Danish Institute for Human Rights, which is a state-funded, independent organisation that has developed a five-step process for conducting HRIAs that has been widely adopted.[37] The HRIA process begins with a planning phase, followed by data collection and assessing impact severity. The final stages are the development of impact mitigation and reporting.

For recommender systems, the DSA has adopted the practice of assessing the possible impact before the deployment of new functionalities and regularly during their life cycle. In addition, the DSA promotes different perspectives, hence allowing audits and assessments to be conducted by multiple actors, from platforms to independent researchers and civil society. Several recitals and provisions in the DSA refer to different kinds of audits and assessments and their targets, be it algorithmic decision-making systems and/or compliance with due diligence requirements. These requirements mainly affect very large online platforms (see details below), but are also relevant to smaller platforms.

Until now, civil society actors, researchers and regulators have already conducted audits and assessments of various online platforms and have found a variety of potential issues and harms posed directly or indirectly by the recommender systems these entities employ. However, in contrast to what is the case in other fields, no specific guidelines for conducting systematic audits of these systems have yet been proposed. Before we present our approach, we want to clarify the terminology.

## 4.1. Audits and assessments in the DSA

The DSA specifies several different audits and assessments for very large online platforms. Some are meant to be conducted by the platforms themselves, while others shall be conducted by contracted, independent auditors. In addition, the DSA mentions other stakeholders who should be able to conduct evaluations regarding algorithmic recommender systems

37 The Danish Institute for Human Rights, 'Human Rights Impact Assessment Guidance and Toolbox', accessed 15 December 2022, https://www.humanrights.dk/tools/human-rights-impact-assessment-guidance-toolbox.

of VLOPs. The following list gives an overview of who should (be able to) evaluate what:

1. *Providers of VLOPs shall carry out risk assessments* – even before deploying functionalities as far as these are likely to have a critical impact on systemic risks (Article 34). If such risks are identified, VLOPs must put mitigation measures in place (Article 35).

2. *Independent organisations/contractors, commissioned and paid by VLOPs, shall audit the compliance of VLOPs* with all due diligence obligations mentioned in Chapter III of the DSA, including the internal risk assessments and risk mitigation measures (Article 37).

3. *Vetted researchers shall be provided access to requested data to conduct research that contributes to the detection, identification, and understanding of systemic risks and the assessment of risk mitigation measures* (Article 40(4)).

4. *Digital Services Coordinators and the European Commission shall have access to data necessary to monitor and assess compliance with the regulation* – including the risk assessments, audits, and mitigation measures (Article 40).

A conventional classification of audit types distinguishes between first-, second- and third-party audits.[38] First-party audits usually are conducted by internal teams; second-party audits are commissioned by platforms themselves but conducted by external contractors or organisations, which are granted access to relevant data; and third-party audits are conducted by independent researchers or entities with no contractual relationship to the audit target. If we want to subsume the audits and assessments required by the DSA under this classification, risk assessments as described in the above list under No. 1 can be understood as first-party audits; independent compliance audits (No. 2) are second-party audits; and independent research by vetted researchers (No. 3) as well as supervision by Digital Services Coordinators and the Commission are third-party audits (see Figure 2).

38 Sasha Costanza-Chock, Inioluwa Deborah Raji, and Joy Buolamwini, 'Who Audits the Auditors? Recommendations from a Field Scan of the Algorithmic Auditing Ecosystem', in 2022 ACM Conference on Fairness, Accountability, and Transparency (FAccT '22: *2022 ACM Conference on Fairness, Accountability, and Transparency*, Seoul Republic of Korea: ACM, 2022), 1571–83, https://doi.org/10.1145/3531146.3533213

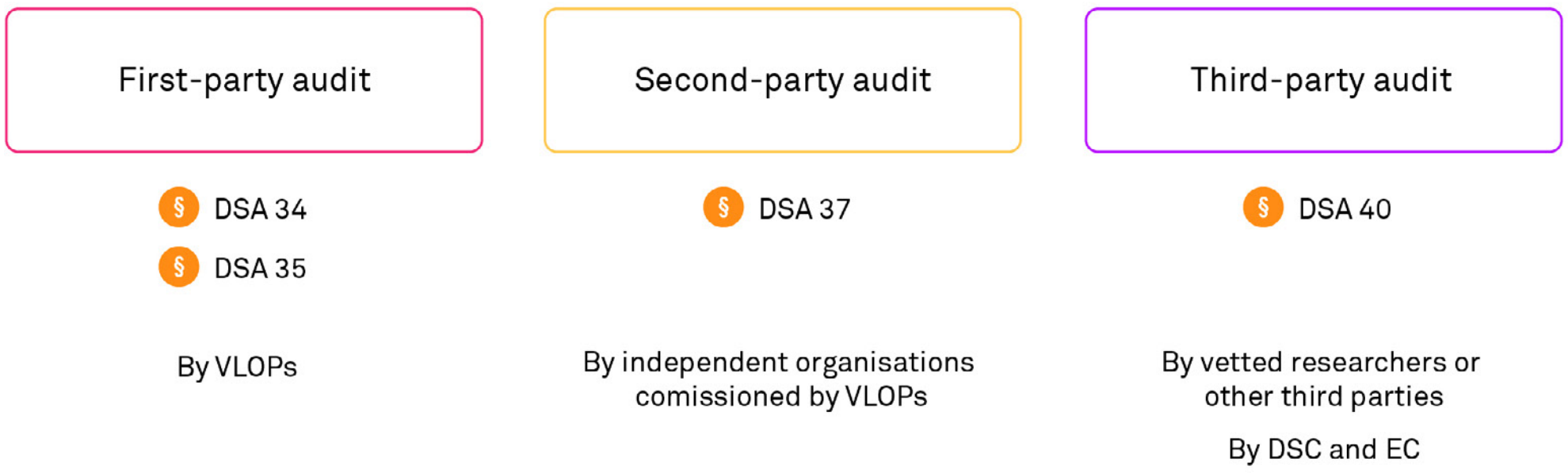

Figure 2: Overview of audit types described in the DSA

Some of these audits and assessments are mandatory and some voluntary, while for others, there will be delegated or implementing acts, and some should be part of a code of conduct. Therefore, a huge variety of multiple perspectives will be taken into account when evaluating systemic risks. Against the backdrop of the DSA coming into force, various stakeholders will be tasked with analysing, assessing and auditing recommender systems.[39] There have been already several proposals for specific methods to analyse social media platforms and systems that use artificial intelligence.[40] This challenge of systematising the fragmented audit and assessment landscape is supposed to be met with voluntary standards (see Article 44, DSA) and voluntary codes of conduct (see Articles 45ff., DSA).

Furthermore—and also relevant to its audit and risk assessment obligations—the DSA describes transparency, data access and reporting requirements in multiple articles. For example, Article 24 details transparency reporting, which should include the number of disputes submitted and number of suspensions. Article 42 adds details for VLOPs, like reporting the number of content moderators, their qualifications and linguistic expertise. Article 27 requires the transparency of recommender systems with respect to the main parameters used for recommendations. Regarding advertising systems, Article 39 describes the details of a repository of advertisements VLOPs need to set up. Article 40 describes the process for VLOPs to give data access to Digital Service Coordinators and the Commission. Additional transparency obligations are requested by civil society to ensure that users can act on the published information.[41]

39 Alexandru Circiumaru and Jenny Brennan, 'Getting under the Hood of Big Tech', 15 March 2022, https://www.adalovelaceinstitute.org/blog/getting-under-the-hood-of-big-tech/.

40 Eticas Consulting, 'Guide to Algorithmic Auditing' (Barcelona: Eticas Consulting, January 2021), https://www.eticasconsulting.com/wp-content/uploads/2021/05/Guide-to-Algorithmic-Auditing-English-Final-ALL-MZ-version7.pdf; Ben Wagner et al., 'Auditing Big Tech: Combating Disinformation with Reliable Transparency' (Tallinn: Enabling Digital, February 2021), https://enabling-digital.eu/wp-content/uploads/2021/02/Auditing_big_tech_Final.pdf; Jenny Brennan, 'Algorithms in Social Media: Realistic Routes to Regulatory Inspection', 3 November 2020, https://www.adalovelaceinstitute.org/blog/algorithms-social-media-realistic-routes-to-regulatory-inspection/.

41 Maximilian Gahntz, 'Towards Responsible Recommending' (Mozilla Foundation, 7 December 2022), https://foundation.mozilla.org/en/research/library/towards-responsible-recommending/towards-responsible-recommending-report/.

## 4.2. Terminology

One challenge when specifying DSA-related audits is that the terms 'audit' and 'assessment' are often used interchangeably in different fields. Still, the understanding of each term can vary widely, as the Ada Lovelace Institute[42] has pointed out. Now, add to this that the DSA takes up these terms, specifies and uses them in its own way. This is why talking about audits and assessments can be quite confusing.

In general, we can understand audits as 'tools for interrogating complex processes, often to determine whether they comply with company policy, industry standards or regulations'.[43] Although in the regulatory arena, including the DSA, an audit is considered to evaluate *compliance* with due diligence obligations, in research, an audit is focused on the identification of a bias within an algorithmic system. These two can overlap, but they do not need to.

*Algorithm* audits are often conducted on a more *technical level* and are part of the regular process of developing machine learning (ML) models. These evaluations are conducted to assess the classification accuracy, computational resources, stability of the algorithm and robustness to specific variations (noise) in the training data. This can also encompass methods for evaluating the potential risks.

These algorithm audits, in turn, can be differentiated from *algorithmic risk and impact assessments*, which are often conducted during or along the early development phase of a new system or ML-based feature to influence its design. After a system is put into production, assessments can serve as a reference to check if or to what extent risks were successfully mitigated. Thus, their focus lies more on the social consequences.

This differentiation between algorithm audits and risk assessments is essential for clearly understanding the scope. Risk and impact assessments often have a wide scope and try to anticipate the societal impact outside of the algorithmic system itself. Algorithm and other system-specific audits, i.e., bias audits, focus on testing a hypothesis that can be evaluated on a dataset or within a system itself. To audit and assess recommender systems, we need to take both perspectives into account because the risks of recommender

42 Ada Lovelace Institute and DataKind UK, 'Examining the Black Box: Tools for Assessing Algorithmic Systems' (London: Ada Lovelace Institute, April 2020), https://www.adalovelaceinstitute.org/wp-content/uploads/2020/04/Ada-Lovelace-Institute-DataKind-UK-Examining-the-Black-Box-Report-2020.pdf.

43 Inioluwa Deborah Raji et al., 'Closing the AI Accountability Gap: Defining an End-to-End Framework for Internal Algorithmic Auditing', 2020, 34, https://arxiv.org/pdf/2001.00973.pdf.

systems are often societal but based on a specific technology. To assess the impact of a recommender system, it must be evaluated along a specific risk scenario for a specific system.

How do we now bring these definitions together with the definitions of the DSA?

As mentioned above, in Article 37, the DSA requires that VLOPs hire external contractors to conduct second-party audits and audit compliance with the VLOPs' due diligence obligations. Although risk assessments are a crucial part of these obligations (see Article 34), it is likely that external auditors will primarily check *if* the VLOPs under scrutiny conduct and document risk assessments (and if necessary, implement mitigation measures), not *how* the risk assessments were carried out and if they were valid and reliable. That is why we will speak of **DSA compliance audits** when referring to audits regarding Article 37 of the DSA.

The socio-technical approach that we propose in this document, however, goes beyond mere compliance, strongly focusing on the auditing and assessing of risks and impacts of algorithmic decision-making systems. This includes several forms of impact and risk assessments, as required in the DSA. Therefore, we distinguish **DSA compliance audits** from **the audit process based on risk scenarios**. For this approach, we follow the five steps of an audit process as defined by the International Standards Organization: planning, risk assessment, audit strategy, evidence gathering and reporting.[44]

The following helps sum up and provide clear terminology throughout the paper:

1. When we refer to Article 37 requirements, we speak of this as a **DSA compliance audit** because it is unclear whether this audit process includes any technical aspects.
2. When referring to any *assessments of risks and impacts*, be it a first-party 'risk assessment' conducted by VLOPs internally (i.e., Article 34, DSA) or a third-party assessment carried out by external and independent researchers (i.e., Article 40, DSA), we write about **risk assessments** and/or impact assessments.[45]
3. When we talk about methods and measurements and how algorithms

44 Jacqui Ayling and Adriane Chapman, 'Putting AI Ethics to Work: Are the Tools Fit for Purpose?', AI and Ethics, 12 September 2021, https://doi.org/10.1007/s43681-021-00084-x, p. 416.

45 Ada Lovelace Institute and DataKind UK, 'Examining the Black Box', therefore, suggest differentiating between 'Algorithm Audits' and 'Algorithmic Impact Assessments'.

can be evaluated *through specific technical approaches*, such as sock puppets, we refer to them as **algorithm audits**. These measurements are not mentioned or specified in the DSA but are necessary for any audit—be it first-, second- or third-party.

Our approach will mainly focus on numbers two and three. Therefore, we see risk assessments and algorithm audits as parts of our general audit process, which we refer to as a **risk-scenario-based audit process**.

# 5. An audit process using risk scenarios

Whether planning a first-, second- or third-party audit, our approach, which uses risk scenarios in a multistakeholder process, will help to structure your audit. Using risk scenarios enables the discussion, operationalisation and prioritisation of the systemic risks, as described in Section 3, and, therefore, facilitates the selection of the most appropriate assessment technique to test those scenarios.

Focusing on systemic risks in concrete **scenarios** will help you clarify questions like the following: What is the definition of 'hate speech' or 'mental well-being'? Where should this risk be observed and how? How does it manifest on a specific platform? What is the impact? Moreover, the process is intended to improve the **observability** of VLOPs and their systemic risks while making mitigation measures more observable. Only when based on a shared understanding of what an *audit process* should look like and how it should work is it possible to compare algorithm audits over time or compare different research approaches regarding the same systemic risks. A common process also helps us structure discussions about measures for transparency, accountability and observability[46] because platforms should be required to offer the possibility of observing processes and, thus, changes. After all, as mentioned at the beginning, any change, no matter how small, to one of the various (sub)algorithms can lead to a momentous shift in the entire recommendation system.

The **risk-scenario-based audit process** can be executed as follows:

1. Find a diverse multistakeholder team.
2. Define and prioritise risk scenarios.
3. Develop, prioritise and conduct measurements.
4. Evaluate measurements and develop a report.

46 Rieder and Hofmann, 'Towards Platform Observability'.

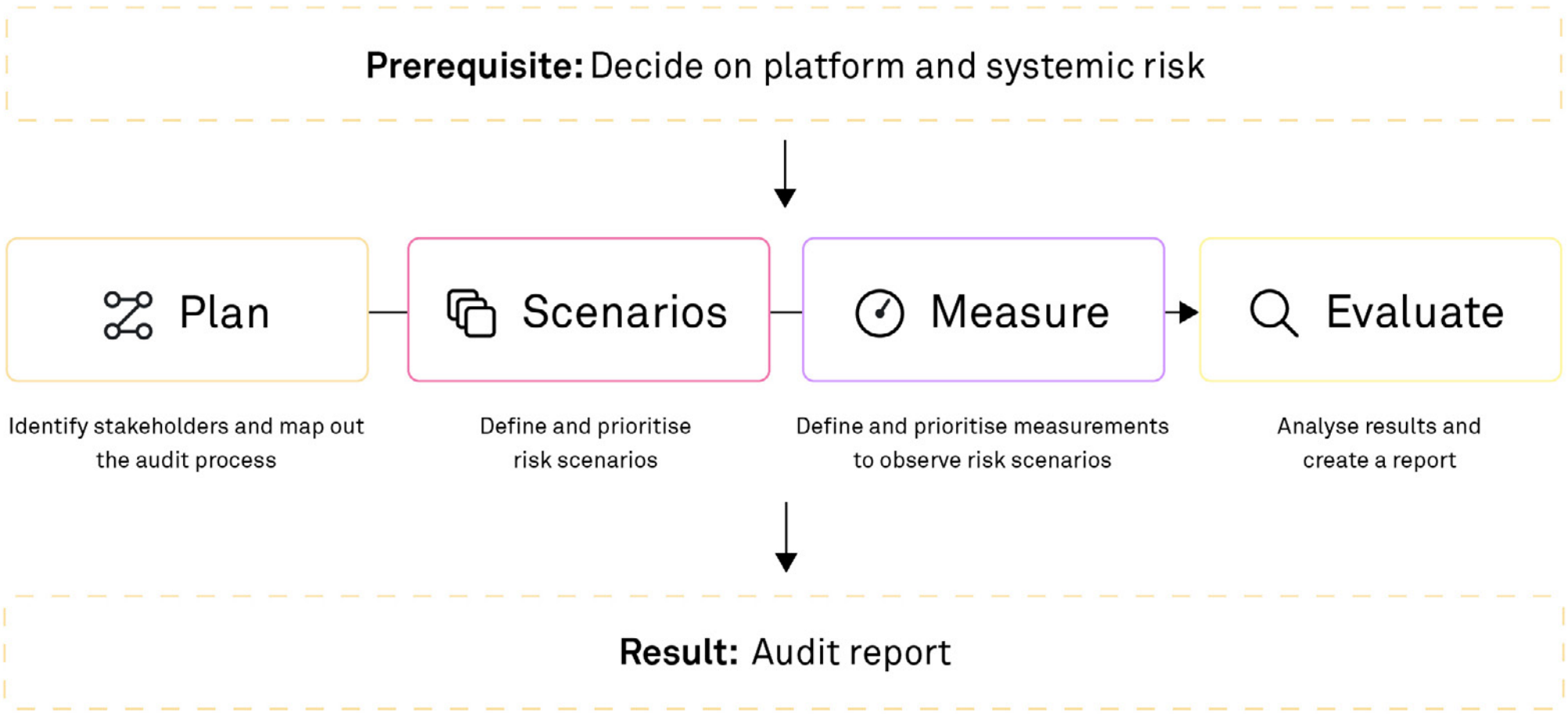

Figure 3: A visual guide to the risk-scenario-based audit process

As described in Section 4.1, audits can be performed by first-, second- and third-party auditors. The process we describe can be used for any of these audit types, but different auditors might focus on different aspects of the process. Because of access restrictions, third-party audits may be limited to methodologies that rely on publicly available data. First-party audits, in contrast, have privileged access to internal documents and data, but the question about the independence of these audits will remain. The same can be true for second-party audits, which are sandwiched between first- and third-party audits and, compared with the latter, may have privileged access to internal documents and data, for example, when conducting compliance audits. Using the risk scenario process in a transparent and documented manner can help reduce questions about independence.

# 6. Step 1: Planning

In the first stage of an audit process, it is necessary to understand the platform that will become the focus of the audit (if that is not already clear, e.g., in an internal audit) and identify those stakeholders that should be involved in the process.

## 6.1. Understand the platform

If you want to find out if platforms cause or at least influence systemic risks, there is no one-size-fits-all solution. As explained in the second section, a platform may consist of several recommender systems, with all of them being embedded in the platform-specific affordances and choice architecture. Therefore, the first step of the audit process is to focus on one platform and try to understand it and its user base.

**Based on that decision, you will have to answer the following questions** to get a basic understanding of the platform and the company behind it:

- **What is the platform's main strategy?** Platforms optimise their recommenders for the companies' goals. For example, The New York Times revealed that TikTok is trying to maximise the number of daily active users and time spent on the app as a means to maximise revenue through advertising.[47] The main strategy also influences the other aspects on this list.

- **What is its audience?** Platforms have different target audiences that make some risks more likely than others; for example, TikTok has a younger user base, increasing the risk of harm to minors.

- **What type of media is the platform based on?** Recommenders for video content are different from those focusing on text; therefore, it is important to obtain a comprehensive understanding of the technologies a platform relies on. For example, TikTok focuses on video, but the aspect of remixing music is an important part of the choice architecture, too.

- **What technical products exist on the platform?** Platforms have different features that are often internally managed by different product teams, meaning they might all behave differently and, therefore, may require separate algorithm audits. For example, TikTok's main recommender is used on the 'For You' feed, but there is also one for the 'Following' page and an additional recommender for search suggestions.

## 6.2. Identify stakeholders

Along with others, Frances Haugen, a Facebook whistle-blower, emphasised that audits and assessments need to be multistakeholder processes that involve civil society and the platforms themselves to be effective and successful.[48] The DSA has also recommended involving different stakeholders in the risk assessment process. We agree that this is indeed indispensable because, as suggested by Helberger et al., 'the realization of public values in platform-based public activities cannot be adequately achieved by allocating responsibility to one central actor (as is currently common practice) but should be the result of dynamic interaction between

47 Kalley Huang, Isabella Simonetti, and Tiffany Hsu, 'TikTok Builds Itself Into an Ads Juggernaut', *The New York Times*, 14 November 2022, https://www.nytimes.com/2022/11/14/technology/tiktok-ads-social-media.html; Smith, 'How TikTok Reads Your Mind'.

48 Frances Haugen, 'Civil Society Must Be Part of the Digital Services Act', *Financial Times*, 29 March 2022, https://www.ft.com/content/99bb6c10-bb09-40c0-bdd9-5b74224a5086.

platforms, users, and public institutions',[49] as well as independent research. Hence, regardless of who sets up the audit process, whether it is a first-, second- or third-party audit or who leads it, you need to include a diverse group of stakeholders.

Although the groups are not exclusive, we identified the following expert groups as important stakeholders who should be involved in a recommender system audit:

1. **Platforms:** To ensure a comprehensive and integral audit and assessment process, it is crucial to include the platforms themselves: the integrity team members, developers or team members of the different products and/or algorithms, as well as developers and experts for data exchange and data interfaces, should ideally all take part. This applies to first-party audits, but it can also be useful for second- and third-party audits to which public policy staff are currently made available but that usually lack the relevant technical expertise. Members of the following platform teams should be considered part of the audit process:

    a. **Platform development:** Those who know how a specific feature is implemented, what data were used to train a ML model and so forth.

    b. **Platform policy:** Those familiar with general management decisions or the platform's values and how they are (ideally) implemented.

    c. **Platform user experience:** Platform employees who know more about how the platform is used.

    d. **Platform research:** Those with expertise on how the platform is used and who may have already conducted studies to understand how mitigations against risks may be implemented.

    e. **Integrity team:** Those working on the research on and mitigation of platform-specific risks, safety and user protection.

2. **Users and civil society:** The perspectives of those affected by systemic risks are crucial to assess. The user perspective can be provided by different actors, such as representatives, actual users/creators or civil society organisations that have experience with the user perspective.

    a. **Representatives of affected groups:** These groups can differ

49 Natali Helberger, Jo Pierson, and Thomas Poell, 'Governing Online Platforms: From Contested to Cooperative Responsibility', *The Information Society* 34, no. 1 (1 January 2018): 1, https://doi.org/10.1080/01972243.2017.1391913.

significantly depending on the risk under investigation. They range from civil society organisations and creator representatives to the affected persons of discriminated groups or child and youth protection representatives. Here, it is essential to think beyond the user base. There might be impacted individuals or communities not using the platforms themselves. Hence, it might be conceivable to create checklists at least for each systemic risk as to whom should be considered.

**b. Users:** Every user can be part of an affected group, so it is necessary to acknowledge user perspectives, especially for more explorative assessments and the detection of risk scenarios, that are not already thought of. Including users will also help to better understand the individual effects of certain risks.

**c. Civil society:** When it comes to the investigation of social media platforms, their risks and their potential impacts on society, civil society organisations have already been working on these topics for years. There are a lot of experts in the field who are working together with academic researchers but also investigating platforms by themselves with explorative methods; these members of civil society organisations enrich the audit process with their perspectives on platforms and risks, often bringing expertise from a variety of fields, such as perspectives of research, affected parties and technological expertise.

**3. Researchers:** Academic and nonacademic researchers have been studying various platforms and their uses in the past. They should be included for their independence and expertise.

**a. Risk researcher:** As described in Section 3, one challenge of risk assessments is to break down abstract systemic risks into testable hypotheses. At this point, researchers and experts on the specific risks under investigation will be relevant. These individuals can be experts from sociology, psychology and related fields of expertise. They should either have dedicated expertise in the risks under scrutiny and their personal, psychological, and/or societal effects and/or be experienced in translating abstract risk categories into concrete scenarios. They need to define, design and discuss hypotheses and approaches to test them. Therefore, it is also important to distinguish among the risks that can be tested on the platform (e.g., via algorithm audits) and more complex societal risks (e.g., via impact assessments).

**b. Audit/assessment researcher:** Researchers who are experienced in risk assessments and audits are the ones who bring technological

expertise to the table. They are the experts in data collection, data evaluation and the fitting between risk scenarios and audit tools. Although these external experts would not be necessary for first-party audits conducted by the VLOPs themselves, they are highly important for external second- and third-party audits.

4. **Other:** Depending on the platform or the systemic risk that is the focus of your audit, experts from other areas should be considered. The following are additional suggestions:

   a. **Legal experts:** Not every audit requires the involvement of legal experts. However, in case the audit should be part of a regulatory inspection, a compliance audit, help with the development or implementation of standards (as, e.g., mentioned in Article 44 of the DSA) or should fulfil any requirements mentioned in the DSA, legal experts need to be part of the process. Depending on the audit's goal, these can be regulators, lawmakers, human rights experts, compliance lawyers or litigators.

   b. **Independent contractors:** Finally, independent contractors are a group of people who should be further specified during the process of implementing the DSA in the near future. In Article 37, the DSA requires external audits for very large online platforms. As explained in Section 4.1, this can also encompass the auditing of internally conducted risk assessments. Therefore, it could be helpful to also involve independent contractors in the assessments and the audits of recommender systems because they bring specific audit expertise from other fields (e.g., finance, due diligence, etc.). In turn, their audits would highly benefit from the expertise of researchers, civil society and representatives of affected groups.

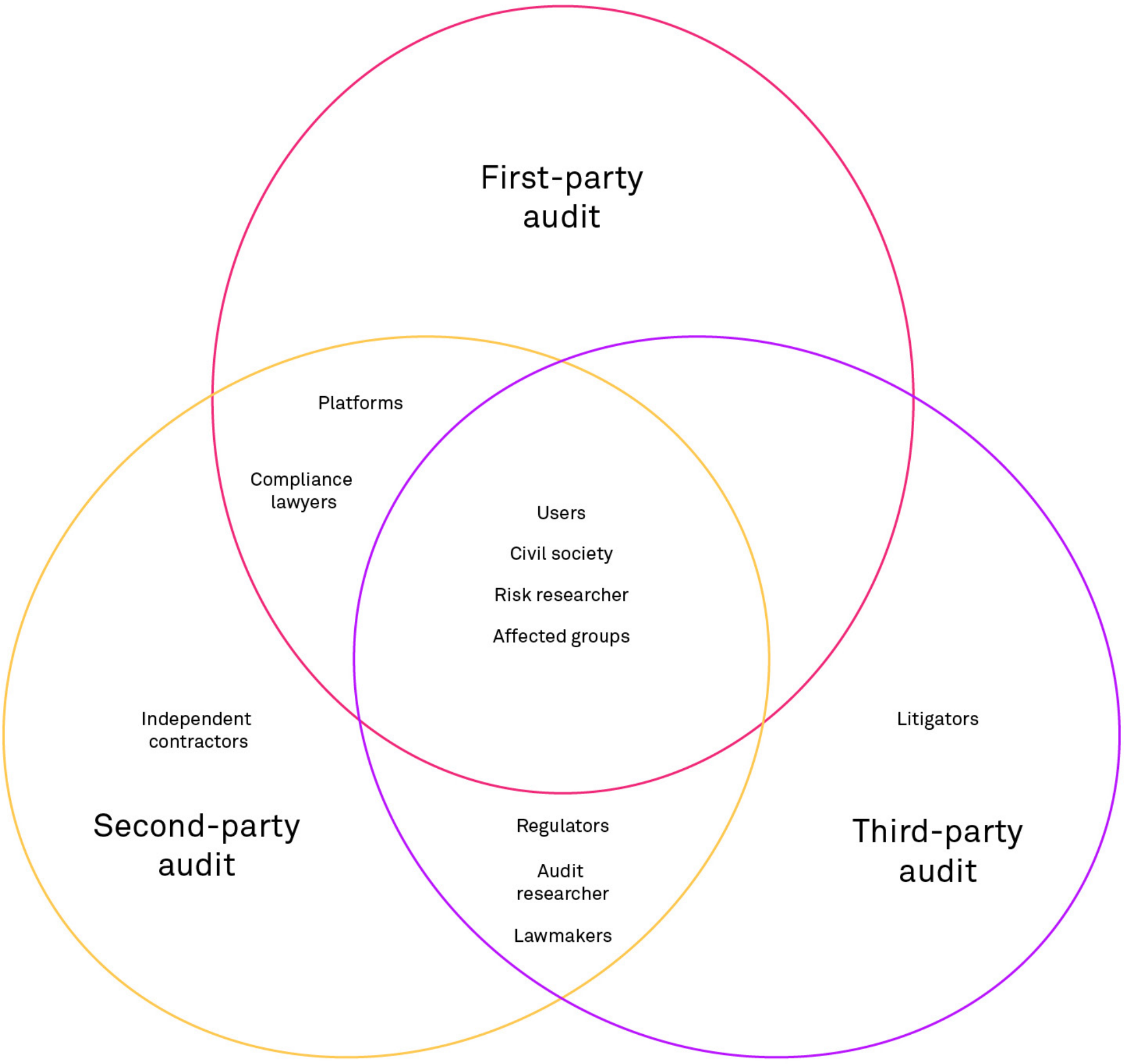

Figure 4:
Audit types and their
suggested participants

Figure 4 provides an overview of which stakeholders should be included in what type of audit. The representatives of affected groups, users, civil society and researchers must be integrated into the process, regardless of who is conducting the audit. In addition, lawmakers, regulators and (external) audit experts should always be part of second- and third-party audits, too. Litigators have a special position in third-party audits because they can contribute their expertise regarding strategic litigations against platforms that negatively affect the risks they are audited for. To ensure the independence of the audit, it must be carefully considered if the representatives of platforms should be involved in the process. Nevertheless, platforms should contribute data and transparency information through public channels.

# 7. Step 2: Define and prioritise risk scenarios

Once the planning process has been completed, the target platform has been defined and the stakeholders identified, the next step is deciding which risk will be the focus of the audit and evaluated concretely.

## 7.1. Scenario definition

Because high-level risks are impossible to assess as a whole, you need to break them down into the potential harms they entail. For example, the question of whether TikTok has serious negative consequences on its users' mental health is hard to answer. It could be helpful to differentiate between users with pre-existing chronic mental health conditions, users with acute mental health issues and users who could be considered unaffected.

Each harm can be further described by a (potentially large) number of scenarios. For example, users in a personal crisis might be more vulnerable when overexposed to specific content (in a 'rabbit hole'). In addition, regardless of the content they are watching, they could be threatened simply by how the recommender works. By fostering prolonged usage, the app prevents them from having contact with friends and family, which could have potential positive health effects. A scenario always describes a hypothetical situation, but depending on what is already known, some aspects might be more realistic or already proven. In contrast, other hypotheses need to be verified through the audit. For the example above, we could back up the vulnerable status of users in a personal crisis through research; here, there are also examples of rabbit holes on TikTok leading to excessive exposure to depressing or negative content as shown by research, *The Wall Street Journal* conducted.[50] However, whether both are connected and increase the risk for the user's mental health would be the subject of the audit.

To standardise the way scenarios are described, we propose including at least the following information:

1. Who is the **affected party**?
2. What **characterises** this party?
3. What **harm** does this actor, person or group experience?
4. How is the **platform** involved in this experience?
5. What are the **macro impacts** that go beyond the individual?

50 WSJ, 'Inside TikTok's Algorithm'.

Answering these questions should allow for a thorough characterisation of the risk scenario. The following descriptions should clarify what is required of each component:

**1. Affected party:** Every risk scenario starts with the definition of the affected party. Who do you think of as being affected by specific harm? Individuals or groups? Recipients or creators? Private persons or advertisers? All platforms have different user groups that show different usage behaviours because there is no 'general user'. There is often a fluid transition between someone consuming content for various reasons and those using a platform for professional and business purposes or creators with different levels of expertise and professionalism. Many platforms also allow organisations, from civil society to companies, to have accounts on their systems.

**2. Characteristics:** The party (potentially) experiencing harm by being on the platform might have characteristics that are relevant to this process: for example, users from marginalised groups, creators with a specific agenda or organisations engaged in a policy debate might be more likely to experience a specific harm. The relevant characteristics should be specified as clearly and thoroughly as possible.

**3. Harm:** This component of the scenario should describe what the affected party experiences that would be considered harmful. Harm is always related to the risk under investigation, to the affected party and its characteristics or traits. Harms can be quite individual and encompass even psychological effects, such as 'feeling depressed', in which harm is connected to mental health risks. If the affected party is a group, for example, a community posting empowering videos on living with mental health issues, harms can have an effect on individual, collective and macro levels. For example, the tightrope of moderating mental health content could lead to overblocking and shadow banning content about mental illnesses, which is meant to empower people. Demoting or blocking that content is a risk that can influence the personal well-being of a user experiencing content about their mental health issues being blocked after sharing it, but this is also an example of discrimination and freedom of speech.

**4. Platform involvement:** As explained in Section 2, several elements and products of a platform can be involved in promoting a specific risk or harm. This part of the scenario should describe or hypothesise what elements of the platform are influencing the recommendation system of a specific product and, therefore, directly or indirectly fostering the harm, such as content moderation (e.g., shadow banning), the user interface (e.g., angry emojis) or user experience (e.g., frictionless endless feeds).

**5. Macro impact:** In Section 3, we noted that many of the 'systemic risks' mentioned in the DSA are abstract. When applied to the risk scenarios, this means that risks can have a negative impact on individuals and groups, as well as on society as a whole. To differentiate between these levels, every risk scenario not only defines a specific harm, but it can also define the larger impact this harm can have on society, for example, civil discourse. These platform effects might be harder to assess.

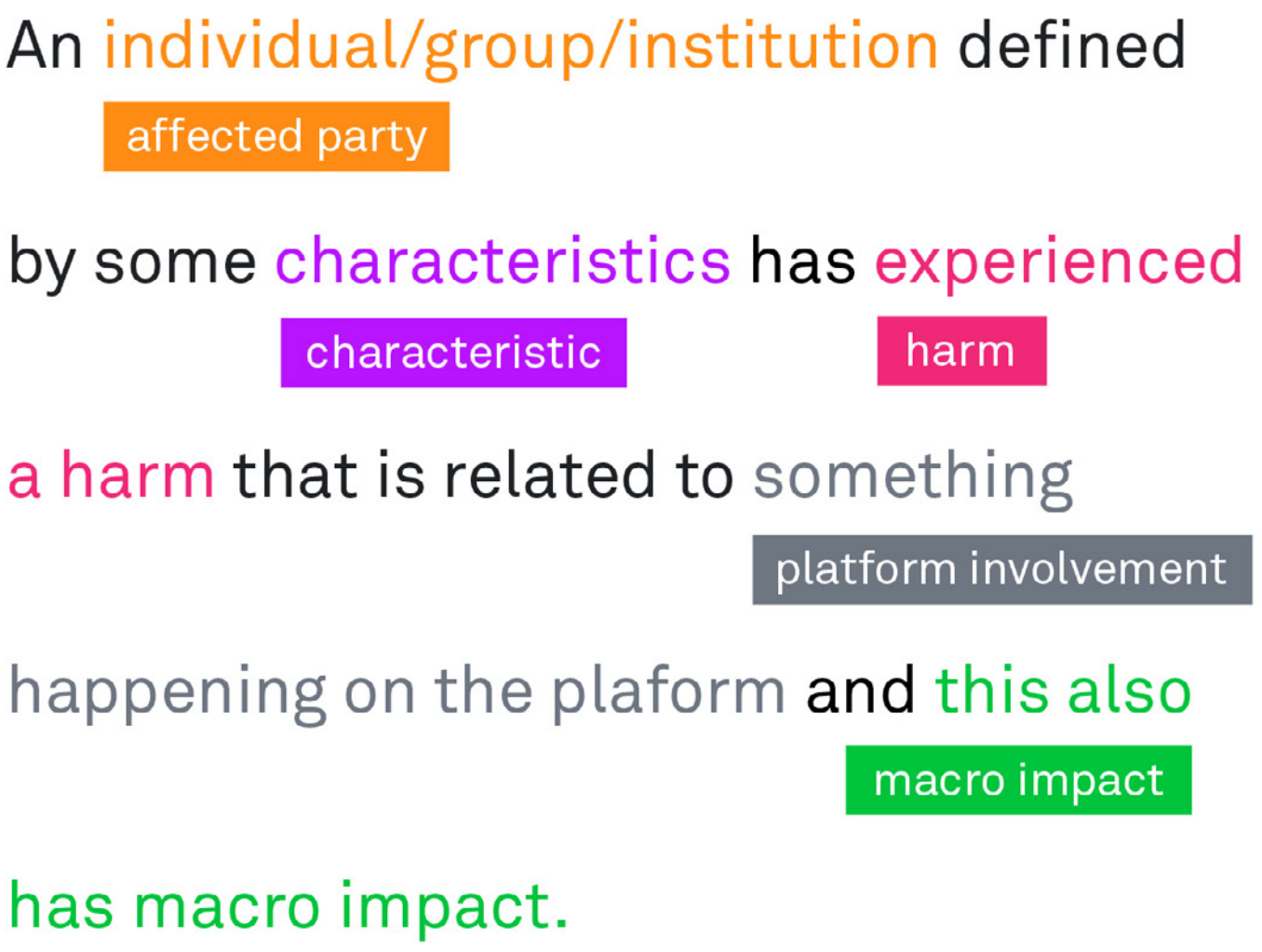

Figure 5: Generic risk scenario description

Putting these components together, a generic risk scenario would look like the following:

To illustrate the scenario definition process, we again refer to an investigation by The Wall Street Journal, which examined TikTok's 'For You' feed.[51] The components of one possible risk scenario derived from that research could be the following:

**Affected party:** Individual user

**Characteristics:** Early 20s, in a personal crisis

**Potential harm:** Excessively exposed to videos describing or showing self-harm

**Platform involvement:** Recommender system of the 'For You' feed

**Macro impact:** Worsen mental health crisis for young adults[52]

51 WSJ.

52 See Matt Richtel and Annie Flanagan, ''It's Life or Death': The Mental Health Crisis Among U.S. Teens', *The New York Times*, 24 April 2022, https://www.nytimes.com/2022/04/23/health/mental-health-crisis-teens.html. The number of emergency room visits by children and adolescents in the US has been increasing since 2009.

The corresponding risk scenario description could then be:

A young adult who is temporarily in a personal crisis is overly exposed to videos describing or showing self-harm by the recommender system of the 'ForYou' feed and this might exacerbate the general mental health crisis of young adults.

affected party · characteristic · harm · platform involvement · macro impact

Figure 6:
Example of a concrete risk scenario description

Other scenarios related to the same systemic risks may involve individuals with the *characteristic* of pre-existing mental health issues suffering the *harm* of societal stigmatisation for being obese or having nonconforming body types where the *platform is involved* because their content is also distributed to users who are known to post comments that are considered bullying. It could cover the *affected party* of activists who post empowering videos for these groups where the *platform involvement* is that they are shadow banned, so the *harm* is that they cannot reach their audience. As another scenario, it could be related to an advertising system (*platform involvement*) that allows businesses to target certain *affected parties* with ads for drugs whose effectiveness has not been assessed by any authority.

## 7.2. Prioritise scenarios

The examples above describe only a subset of scenarios for one systemic risk. Therefore, even if you focus on only one risk or have an interest in implementing a specific audit, you might find yourself with more risk scenarios than is possible or advisable to evaluate.[53] Because of this and given that resources of every kind are always scarce, prioritising becomes paramount. Establishing priorities allows you to decide which scenarios

53 In the future, we would like to create an inventory of risk scenarios to be shared with the community that could serve as a starting point for other types of impact assessments.

should be tackled first. The question then becomes which criteria should be used to rank the scenarios.

> 'In determining the significance of potential negative effects and impacts, providers should consider the severity of the potential impact and the probability of all such systemic risks. For example, they could assess whether the potential negative impact can affect a large number of persons, its potential irreversibility, or how difficult it is to remedy and restore the situation prevailing prior to the potential impact.' (Recital 79, DSA)
>
> 'This risk assessment shall be specific to their services and proportionate to the systemic risks, taking into consideration their severity and probability [...].' (Article 34(1), DSA)

From a legal perspective, the DSA specifies that auditors should estimate the probability and severity of systemic risks. Using occurrence probability and impact severity as criteria to evaluate and prioritise risks is quite common in many audit paradigms. In some applications, risk is even operationalised as the arithmetic product of these two factors. Others have used more nuanced metrics. For example, the GDPR suggests using 'likelihood' instead of mathematical 'probability' for DPIAs. However, we argue that it is not expedient to mathematically calculate these risks. Estimating the 'impact' of human rights violations, as sensitive as the deterioration of an individual's self-esteem, by assigning a number could be considered too subjective and even borderline cynical. Similarly, objectively determining the probability or likelihood that these harms will take place is simply impossible. Instead, we propose to rank scenarios in groups of **'high'** and **'normal'** priority by considering the factors mentioned in the recital above (number of affected persons, irreversibility of the harm and potential for remedies) and whether the platform already has mitigation measures in place.

We only suggest these two ranking categories because the scenarios under consideration, such as threats to fundamental rights, should never have low priority. If the spectrum of scenarios under consideration is too wide, you can also add a very high-priority category.

Imagine the following scenario: A young adult with mental health issues becomes poisoned after ingesting chemicals advertised as pseudo-medications because the recommender system has amplified this misinformation within a cluster of users.

Although the actual number of people who harm themselves might be low and there is also a responsibility on behalf of the users, the irreversibility of the harm is severe; therefore, the scenario is labelled a high priority.

Now, think of users who feel insecure about their bodies but do not suffer from any mental health issues. They could be targeted with advertisements that perpetuate unhealthy physical appearance stereotypes, pressuring them to conform to these norms. Although there might be grave effects for individuals continuously confronted with the unachievable appearance norms in advertising, this potential harm stems from often sexist cultural norms affecting societies at large, not the platform or the recommender system. Hence, the risk scenario can be labelled as a normal priority because the risk it poses is not any different than those that users are exposed to when interacting with other media outlets. Suppose the audited platform focuses on fashion or adult content. In this case, this same instance may very well be labelled as high priority because the whole platform focuses on this content, not just advertisements.

Once the risk scenarios have been prioritised, you must decide which scenarios should be further investigated and what exactly to measure. Some factors to consider here are financial resource availability, technical expertise at hand and organisational focus and anything that may constrain the development of the audit.

# 8. Step 3: Developing measurements

After developing and prioritising scenarios, we need to achieve an understanding of how they can be observed on the platform. As described above, there is no generic social media platform. Instead, each has its focus (e.g., type of media), multiple products (e.g., search recommendations and what to consume next) and affordances, which means it supports different kinds of interactions. These affordances are reflected in the elements of a platform that form a choice architecture guiding users through the system. Moreover, each platform has an architecture comprised of different internal systems. Therefore, how a scenario can be observed—and, hence, measured—depends on the specific platform. The methods will differ depending on the platform design and available audit resources.

In the following two subsections, we will explain, first, a list of types of audits and, second, platform elements that should be considered when developing measurements for a specific product of a platform. Figure 7 summarises these lists and shows the connections between types of audits and elements of the platform.

Figure 7:
Overview of types of audits, elements of the platform and examples of which type of audit can produce data on which platform elements.

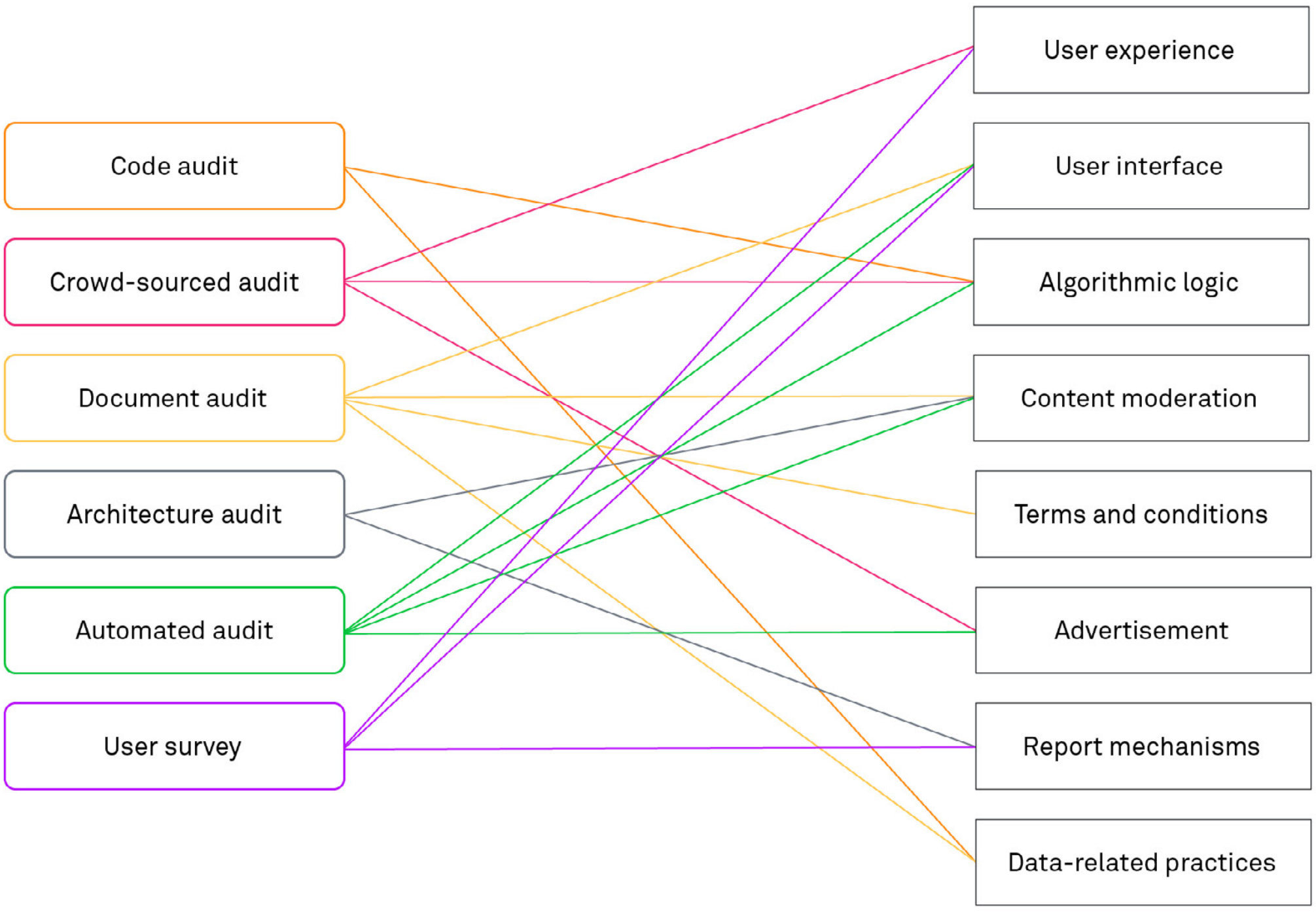

## 8.1. Types of algorithm audits

With reference to the papers by Sandvig et al.[54] and the Ada Lovelace Institute[55], the list below provides an overview of the different kinds of algorithm audits, their benefits, limitations and a few examples. Some types of audits are based on the information they are analysing and some on how data are collected. Auditing a scenario will often require including both aspects.

**Code/data audit:** Given access to the code base, existing user data, ML models and training data, an auditor could review the inner workings of a specific recommender system. That auditor can evaluate the choices of

54 Christian Sandvig et al., 'Auditing Algorithms: Research Methods for Detecting Discrimination on Internet Platforms', in *Data and Discrimination: Converting Critical Concerns into Productive Inquiry*, 2014, http://websites.umich.edu/~csandvig/research/Auditing%20Algorithms%20--%20Sandvig%20--%20ICA%202014%20Data%20and%20Discrimination%20Preconference.pdf.

55 Ada Lovelace Institute, 'Technical Methods for Regulatory Inspection of Algorithmic Systems' (London: Ada Lovelace Institute, 9 December 2021), https://www.adalovelaceinstitute.org/report/technical-methods-regulatory-inspection/.

algorithms and code libraries, ensure that state-of-the-art technology is used or review the ML models, how they were trained and other existing data to check for biases.

*Benefits:* A code audit can help to identify code-related flaws, for example, failure to process certain data elements or biases stemming from the selection and parsing of input data.

*Limitations:* Code audits are challenging to conduct because code is often considered a trade secret subject to intellectual property constraints and the underlying codebase is continuously evolving in short development cycles. Conducting an audit on code alone will not allow auditors to assess systemic risks related to how a platform is used.

*Example:* Wilson et al.[56] conducted an internal audit at Pymetrics, a human resources company using AI to assess job candidates, and evaluated the biases in their screening system. The researchers analysed the algorithmic fairness of Pymetrics' approach regarding its assumptions, correctness and other factors. The study was later criticised for a conflict of interest because the researchers received funding from the company and limited the scope of the audit to technical aspects.[57]

**Crowd-sourced audit:** The users of the platform donate data to share what is shown to them on the platform. They can do so by providing auditors with access to account-related data, for example, through specialised plugins or by copying data following a script provided to them by researchers.

*Benefits:* Data donations can shed light on specific problems experienced on the user side. Compared with automated audits, they provide data from real user behaviour.

*Limitations:* It is very hard to balance the set of participants to get a representative sample because, often, only a specific user group is reached. Users may violate the terms of service of the platform, risking being banned. User data come with high requirements regarding the protection of personal data. It is also often impossible to collect all the necessary data to understand how user behaviour and data provided affect the recommender, for example,

56 Christo Wilson et al., 'Building and Auditing Fair Algorithms: A Case Study in Candidate Screening', in *Proceedings of the 2021 ACM Conference on Fairness, Accountability, and Transparency* (FAccT '21: 2021 ACM Conference on Fairness, Accountability, and Transparency, Virtual Event Canada: ACM, 2021), 666–77, https://doi.org/10.1145/3442188.3445928.

57 Meg Young, Michael Katell, and P.M. Krafft, 'Confronting Power and Corporate Capture at the FAccT Conference', in *2022 ACM Conference on Fairness, Accountability, and Transparency* (FAccT '22: 2022 ACM Conference on Fairness, Accountability, and Transparency, Seoul Republic of Korea: ACM, 2022), 1375–86, https://doi.org/10.1145/3531146.3533194.

the complete watchlist of a years-old account or all comments posted in the past.

*Example:* DataSkop[58] is a data donation platform developed by AlgorithmWatch and several project partners. Open-source data donation software can be used for noncommercial research, civil society projects, and journalistic research. In a pilot project, they gathered data on YouTube recommendations regarding the 2021 German federal election.[59] The participants were recruited through public channels, resulting in biased submissions because a specific user base was more easily reached.[60]

**Document audit:** Many platforms release a broad set of information that can be analysed in multiple ways, for example, transparency reports or terms of service. Sometimes, whistle-blowers publish internal documents that disclose information on how the platform operates. In addition, internal auditors may get access to even more reports produced for internal processes.

This approach requires either cooperation with the platform (which might lead to a conflict of interest), whistle-blowers leaking internal data or collecting other data publications, for example, in court cases. When working with original data, auditors must carefully redact the documents.

*Benefits:* Information is (publicly) available, and data are often aggregated regarding specific questions.

*Limitations:* A limited set of information that may not be relevant for the audit process. Transparency reports are often not comparable between platforms or across multiple years because the methods for aggregating the data are not disclosed and there are no standards for reports. Leaked information is often hard to verify, or it might disclose a procedure for which the current status is unknown.

*Example:* A recent example of a document audit is the so-called 'Twitter Files' published by journalists of 'The Free Press'.[61] Their analysis was based on an examination of the internal communications archive of the company. The results shed light on the internal content moderation processes of the platform and related 'shadow banning' practices. In the case of the 'Twitter

58 'Dataskop', DataSkop, 2022, https://dataskop.net/overview-in-english/.
59 AlgorithmWatch, 'DataSkop: Investigating YouTube's Algorithm during Germany's Election Campaign', *AlgorithmWatch*, 15 July 2021, https://algorithmwatch.org/en/dataskop-investigating-youtubes-algorithm-during-germanys-election-campaign/.
60 Other crowdsourced audit initiatives are for example the 'Citizen Browser' by The Markup, the addons provided by 'Tracking Exposed', the Initiative 'Who Targets Me' or Mozillas 'Regrets Reporter'.
61 Bari Weiss, 'Our Reporting at Twitter', *The Free Press*, 15 December 2022, https://www.thefp.com/p/why-we-went-to-twitter.

Files', the publication of personal data of low-ranking employees resulted in online harassment.[62]

**Architecture audit:** Platforms employ several different systems, including recommenders, and their introduction or interplay may increase or decrease certain risks. Knowing what systems are in use, how they are connected and how data flow between them may help in understanding their combined effects.

*Benefits:* New elements to the architecture can be audited before their introduction, and their effects can be studied on real data during the deployment process.

*Limitations:* The audits of changes in the architecture depend on knowledge that, in most cases, only the platforms have (e.g., what new features are tested on whom). If platforms are transparent about their updates, researchers could compare data from the time before the deployment of new features to data gathered after this change. One of the limitations is that not just the architecture changes, but many factors could influence user behaviour at different points in time, making attributions of causality and, thus, comparisons difficult.

*Example:* Huszár et al.[63] (conducted by Twitter) analysed the impact of the algorithmically personalised 'Home Timeline' of Twitter introduced in 2016 by excluding a control group from the new feature. They then compared the two groups and found that, in the group with the algorithmically sorted timeline, mainstream right-wing posts received higher algorithmic amplification than mainstream left-wing posts. Although this study produced reliable insights on which content is amplified, it did not offer any insights into *why* it is being amplified.[64]

**Automated audit:** Automated measurements through sock puppets, APIs or scraping have been widely used and can be conducted by independent auditors.

62 Isaac Saul, 'The Twitter Files on Hunter Biden', *Tangle* (blog), 5 December 2022, https://www.readtangle.com/twitter-files-elon-musk-hunter-biden/.

63 Ferenc Huszár et al., 'Algorithmic Amplification of Politics on Twitter' (Palo Alto, CA: Twitter, 21 October 2021), https://cdn.cms-twdigitalassets.com/content/dam/blog-twitter/official/en_us/company/2021/rml/Algorithmic-Amplification-of-Politics-on-Twitter.pdf.

64 Following these results, Brown et al. conducted a scraping audit and found that the reason may be that more people interact with mainstream political right posts. Hence, outrage and mocking seem to contribute to algorithmic promotion. This illustrates the importance of combining different audit types because of the limitations of each type. See Megan A. Brown, Jonathan Nagler, and Joshua Tucker, 'Twitter Amplifies Conservative Politicians. Is It Because Users Mock Them?' *Washington Post*, 27 October 2021, https://www.washingtonpost.com/outlook/2021/10/27/twitter-amplifies-conservative-politicians/.

- **Sock puppet audit:** Sock puppets[65] are user profiles created and programmed to imitate human behaviour. They can be untrained, for example, as new, blank accounts, or trained on certain user profiles (e.g., different ideologies or interests), usually by letting them automatically consume or interact with different categories of content.[66]

  *Benefits:* Sock puppet audits can be carried out without platform cooperation. The behaviour of the puppets can be controlled to experiment with the recommender.

  *Limitations:* Even well-trained sock puppets differ from accounts of real humans because auditors must make certain assumptions about user behaviour. Some recommender systems might even be able to detect sock puppets and then act differently.[67] Most platforms also forbid the use of sock puppets in their terms of service.

  *Example:* Haroon et al.[68] trained 100,000 YouTube sock puppets on a specific ideology ('left, centre-left, centre, centre-right and right') by letting them watch 100 randomly sampled videos from their assigned ideology. The study concluded that 'YouTube's recommendations do direct users, especially right-leaning users, to ideologically biased and increasingly radical content'.[69] Although the number of sock puppets in this study was impressively high and the authors made efforts to diversify the simulated profiles, this significantly increases the resources needed and still might not represent actual users.

- **API audit:** In API audits, auditors access platform data via an 'application programming interface' (API). The platforms themselves provide these APIs.

  *Benefits:* APIs give auditors easy access to platform data simplifying data collection and analysis.

  *Limitations:* API audits strongly depend on the platforms and their decisions about what data and how much of it they share through their API. Although some APIs are open to everyone, others require registration; therefore, platforms can restrict access. The data received

65 The terminology differs, common synonyms are 'virtual agents' or 'bots'.
66 Muhammad Haroon et al., 'YouTube, The Great Radicalizer? Auditing and Mitigating Ideological Biases in YouTube Recommendations', ArXiv:2203.10666 [Cs], 24 March 2022, http://arxiv.org/abs/2203.10666.
67 Martin Degeling in: Nicolas Kayser-Bril, 'How Researchers Are Upping Their Game to Audit Recommender Systems', *AlgorithmWatch* (blog), 2 November 2022, https://algorithmwatch.org/en/researchers-audit-recommender-systems/.
68 Haroon et al., 'YouTube, The Great Radicalizer?'
69 Haroon et al., 1.

from an API can also differ from the experiences of real users. Either auditors trust the API or must compare the output to other data (e.g., from crowd-sourced or sock puppet audits).

*Example:* Dhiraj Murthy[70] explored the role of YouTube's recommender system in directing users to ISIS-related extremist content. He queried the YouTube API with 'YouTube Data Tools'.[71] With the API, it was possible to point out whether ISIS videos were recommended to users via the search function or the recommendations beyond a seed video. Murthy not only found that it was possible to find ISIS videos in 2016 (although it was not easy), he also was able to track that, in 2020, YouTube appeared to have made efforts to combat ISIS videos. However, what cannot be assessed with this approach is whether users with existing profiles would get the same recommendations as extracted by the API.

- **Scraping audit:** Scraping audits involve monitoring and processing a platform to collect data, for example, creating a data set of users active on the platform. Scraping is mostly passive when it comes to interaction with the platform.

*Benefits:* Scraping audits do not require the platform's cooperation either and can help get an overview of the platform's content over time, for example, in continuously monitoring newly added content or search results.

*Limitations:* Like sock puppets, scraping often violates the platforms' terms of service.[72] Although this audit focuses on monitoring a service, it does not include personalisation for specific user accounts, which is an essential factor in studying recommenders.

*Example:* Zhou et al.[73] collected data about videos on YouTube, finding that the recommended video section was the most important source of views for the majority of videos. They found strong correlations between the view count of a video and average view count of its top referrer videos, as well as between the position of a video on the recommended list and the click-through rate of the video. Although the data provided insights into the correlations of recommendations and

70 Dhiraj Murthy, 'Evaluating Platform Accountability: Terrorist Content on YouTube', *American Behavioral Scientist* 65, no. 6 (1 May 2021): 800–824, https://doi.org/10.1177/0002764221989774.
71 Bernhard Rieder, 'YouTube Data Tools', 2015, https://tools.digitalmethods.net/netvizz/youtube/.
72 Third-party auditors should check the legal situation in their country before conducting a scraping or sock-puppet audit.
73 Renjie Zhou, Samamon Khemmarat, and Lixin Gao, 'The Impact of YouTube Recommendation System on Video Views', in *Proceedings of the 10th ACM SIGCOMM Conference on Internet Measurement*, IMC '10 (New York, NY, USA: Association for Computing Machinery, 2010), 404–10, https://doi.org/10.1145/1879141.1879193.

view counts, similar to using untrained sock puppets, it could not offer insights into those aspects impacted by personalisation.

**User survey:** Surveying users of a platform can shed light on a broad set of problems experienced by participants. These surveys can follow different qualitative and quantitative approaches.

*Benefits:* Studies with actual users allow auditors to analyse the socio-technical environment. To assess the risks of algorithmic systems, not only the systems themselves matter, but also the experiences with the systems.

*Limitations:* The study design must carefully differentiate between the causalities of issues and the platforms' influence. User surveys usually provide only an overview of users' assumptions; they do not provide any causal information about the algorithm itself.

*Example:* Klug et al.[74] conducted qualitative interviews with TikTok creators on their assumptions about TikTok's recommendation algorithm. Video engagement, posting time, adding trending and algorithm-related hashtags and piling up hashtags were the main metrics creators assumed had an influence on TikTok's recommendation algorithms. Because user surveys can only ask for subjective experiences, the researchers tested these assumptions with scraped data from TikTok videos. This example shows that user surveys can be used to collect assumptions and hypotheses that must be checked with another audit type.

The list we have provided is not exhaustive. There are other audit types as well as other benefits and limitations of each audit. Different audit types can and should be combined to comprehensively audit a scenario.

## 8.2. Elements of the platform

Which audits to use will depend on the choice architecture elements under scrutiny. Each product of a platform has different elements that can increase or decrease risks, and each element's influence needs to be considered and measured. For example, limiting the spread of illegal content depends on what aspects are considered illegal in terms of service and according to (local) law, whether there is an automated detection algorithm at work, how the content moderation rules for this type of content are defined and how

74 Daniel Klug et al., 'Trick and Please. A Mixed-Method Study On User Assumptions About the TikTok Algorithm', in *13th ACM Web Science Conference 2021*, WebSci '21 (New York, NY: Association for Computing Machinery, 2021), 84–92, https://doi.org/10.1145/3447535.3462512.

the interface is designed for users to report such content.

Therefore, 'platform involvement' is part of the risk scenarios we explained above. Consequently, developing measurements to test scenarios requires the use of algorithm audits on a specific platform element. The below list incorporates and extends the platform elements mentioned in the DSA (e.g., Article 34(2)) and should help you sharpen your approach and find the best-fitting algorithm audit. The decision regarding proper measurements for each scenario should be made alongside the algorithm audit types and platform elements. This will help you discuss the benefits and limitations of each approach while focusing on a hypothesis that can be built in conjunction with the scenario. Not all types of audits fit all elements of a platform. Figure 7 provides examples of the most relevant connections.

**User experience/journey:** The user experience or user journey encompasses the subjective experience of a user on a platform regarding its utility and ease of use or efficiency concerning their individual usage. In terms of the user experience, you should ask yourself the following: How do the general aspects of the platform design impact what users can and cannot do on the platform? This aspect includes the overall goal of a platform implemented in its focus on specific content and interaction forms.

*Example:* The starting point of the TikTok App is always the 'For You' feed rather than the 'Following' feed. On YouTube, the autoplay function that starts the next video when the currently viewed one is finished is enabled by default. These elements emphasise the recommended content over content that is, for example, posted by accounts being followed.

**User interface:** The user can interact with the system via the user interface. This includes the control mechanisms made available through the layout and design of, for example, an app or website. Questions to analyse the user interface for an audit are as follows: How does the interface support user interactions that lead to or may help mitigate a risk scenario? This element can range from the number of clicks necessary to perform an action to the language used to encourage or restrain certain actions. You should also consider that the DSA prohibits the use of 'dark patterns' in design that prevent users from making 'free and informed decisions' (Article 25, DSA). Examples of dark patterns are pre-checked boxes (e.g., for newsletter sign-up during online purchases) or cookie consent notices where the 'accept all' button is bigger or more colourful than the 'reject all' button (if present at all).

*Example:* Twitter implemented the 'Retweet' button in 2009, and to this day, it remains one of the key features of the platform. At the same time, the developer regrets that he implemented it because it may have decisively contributed to the emergence of online mobs that spread disinformation and moral outrage.[75] Thus, the design of user interfaces can trigger specific actions by users and may contribute to emerging harms.

**Algorithmic logic:** Algorithmic logic can refer to different products of the platform, such as the following:

- Personalised user feeds (e.g., Twitter Home feed, TikTok 'For You' feed)
- Recommendations based on search queries
- 'Watch next' recommendations (e.g., YouTube 'Watch next')
- Start page recommendations (e.g., YouTube main page)

These examples demonstrate that there is not 'one' algorithmic logic to consider (see Section 2.2), but there are different systems that are partly intertwined and that sometimes work under differing logics. Questions to consider when thinking about measurements for algorithmic logic: How do the automated elements of the platform impact the scenario? What data are used and where? How are content and users categorised and clustered? How do they automatically amplify, demote or ban content?

*Example:* Platforms in general might be optimised for a number of different attributes, for example, daily active users and watch time in the case of TikTok's 'For You' feed. Other platforms or elements might have different optimisation goals; for example, the 'watch next' feature has the goal of keeping users engaged within a session, and a search function might be optimised for ad display.

**Moderation:** Moderation not only includes the removal of content or suspension of accounts, but also a wide range of other measures, such as labelling content, reducing the visibility of content or disabling comments. Besides this, measures can be taken not only against pieces of content, but also against specific accounts.[76] To consider aspects of moderation is especially relevant regarding the risks to freedom of expression or media freedom. Considering moderation for an audit means answering the following question: How do the platform's decisions on moderation and moderation techniques affect the scenario?

*Example:* The 'Twitter Files' triggered a debate on content moderation

75 Alex Kantrowitz, 'Man Who Built The Retweet: 'We Handed A Loaded Weapon To 4-Year-Olds'', *BuzzFeed News*, 23 July 2019, https://www.buzzfeednews.com/article/alexkantrowitz/how-the-retweet-ruined-the-internet.

76 Eric Goldman, 'Content Moderation Remedies', *Michigan Technology Law Review* 28, no. 1 (1 December 2021): 1–60, https://doi.org/10.36645/mtlr.28.1.content.

practices and their opacity. The decisions on suspending accounts and reducing the visibility of certain content were not based on a clear, transparent and comprehensible process.[77] Also, TikTok users have reported that it is difficult to contact the content moderation team, and many bans of user accounts do not list the reasons for the ban.[78]

**Terms and conditions:** Terms and conditions and similar documents, such as community guidelines, describe what content or user behaviour is allowed and accepted on a platform. Only if something is acknowledged as a risk or mentioned as forbidden should actions by the platform be expected. What do the terms of service or community guidelines say about content and risks?

*Example:* At the beginning of the Covid-19 pandemic, it took platforms some time to adjust their rules and prevent the spread of misinformation about the public health crisis. For instance, TikTok did not mention misinformation explicitly before January 2020.[79]

**Advertisement:** Given that social media platforms often have different systems and rules for user content and advertising, some risks are explicitly related to ads. The DSA mentions that these risks can range from advertising illegal or harmful content to discriminatory portrayal in advertising, which can have a negative impact on equal treatment. You should ask yourself the following: Is something specifically related to the advertising elements of the platform?

*Example:* Facebook has been criticised multiple times for allowing political ads to micro-target users, for example, with campaigns that were intended to discourage people from voting. TikTok banned political advertising completely, but researchers have shown that political ads still exist.[80]

**Reporting Mechanisms:** Most platforms allow users to report content they see for various reasons, such as community guideline violations. There are overlaps between content moderation and user interface elements, but the process is also important. To develop measurements, you should ask yourself the following: How does the process of reporting content work? Does it enable user reporting, or does it put up barriers in the scenario? Do those who are reported get any feedback or information about the reasoning?

77 Weiss, ‘Our Reporting at Twitter’.
78 Amelia Som, TOCA: TikTok creators unite against opaque rules, interview by Sebastian Meineck, 25 July 2022, https://netzpolitik.org/2022/toca-tiktok-creators-unite-against-opaque-rules/.
79 Andrew Hutchinson, ‘TikTok Adds New Rules to Ban Harmful Misinformation in the App’, *Social Media Today* (blog), 9 January 2020, https://www.socialmediatoday.com/news/tiktok-adds-new-rules-to-ban-harmful-misinformation-in-the-app/570064/.
80 Mozilla, ‘These Are ‘Not’ Political Ads. How Partisan Influencers Are Evading TikTok’s Weak Political Ad Policies’ (Mozilla Foundation, June 2021), https://foundation.mozilla.org/en/campaigns/tiktok-political-ads/.

*Example:* On TikTok, the feature of reporting content for violations of guidelines is part of the 'share' button, a location where users might not expect it to be. On Instagram, it is highlighted in red and is part of the general content item menu. Both platforms offer a 'Violates NetzDG' item alongside categories of possible violations (e.g., 'mobbing').[81] Most users likely do not know what the NetzDG is, potentially hindering the usage of this reporting mechanism.

**Data-related practices (i.e., data collection/processing):** Platforms process and collect different types of data, from content posted by users to tracking data from interactions. These data are then used and enriched in processes relevant for audits. As an auditor, you should answer the following questions: What role do specific data points or analyses play in this scenario? Which data are being collected and processed (e.g., aggregated or inferred)? How and in what ways are data protection measures implemented?

*Example:* Facebook allowed third-party developers to access users' personal information via third-party apps. This was also used by Cambridge Analytica, a former British consulting firm. Cambridge Analytica closed down after being involved in a data privacy violation scandal for microtargeting political advertisements in the 2016 US presidential campaigns, leading to accusations of influencing the election and having a negative effect on the electoral process.

This list can be extended if necessary.

## 8.3. Prioritisation of measurements

Based on the lists of types of audits and elements of the platform, you and your team will be able to develop several measurements for each scenario that you want to audit. In many cases, the number of possible measurements will exceed the available resources of time, computing power or expertise for a specific type of audit. The following criteria (see Figure 8 below) can help you prioritise the measurements and decide which audit to conduct and in what order.

81 Instagram, 'Network Enforcement Act ('NetzDG') | Instagram Help Centre', accessed 20 December 2022, https://help.instagram.com/130785144276082; TikTok, 'Netzwerkdurchsetzungsgesetz | TikTok Transparency', TikTok, 8 November 2022, https://www.tiktok.com/transparency/de-de/netzdg-about/.

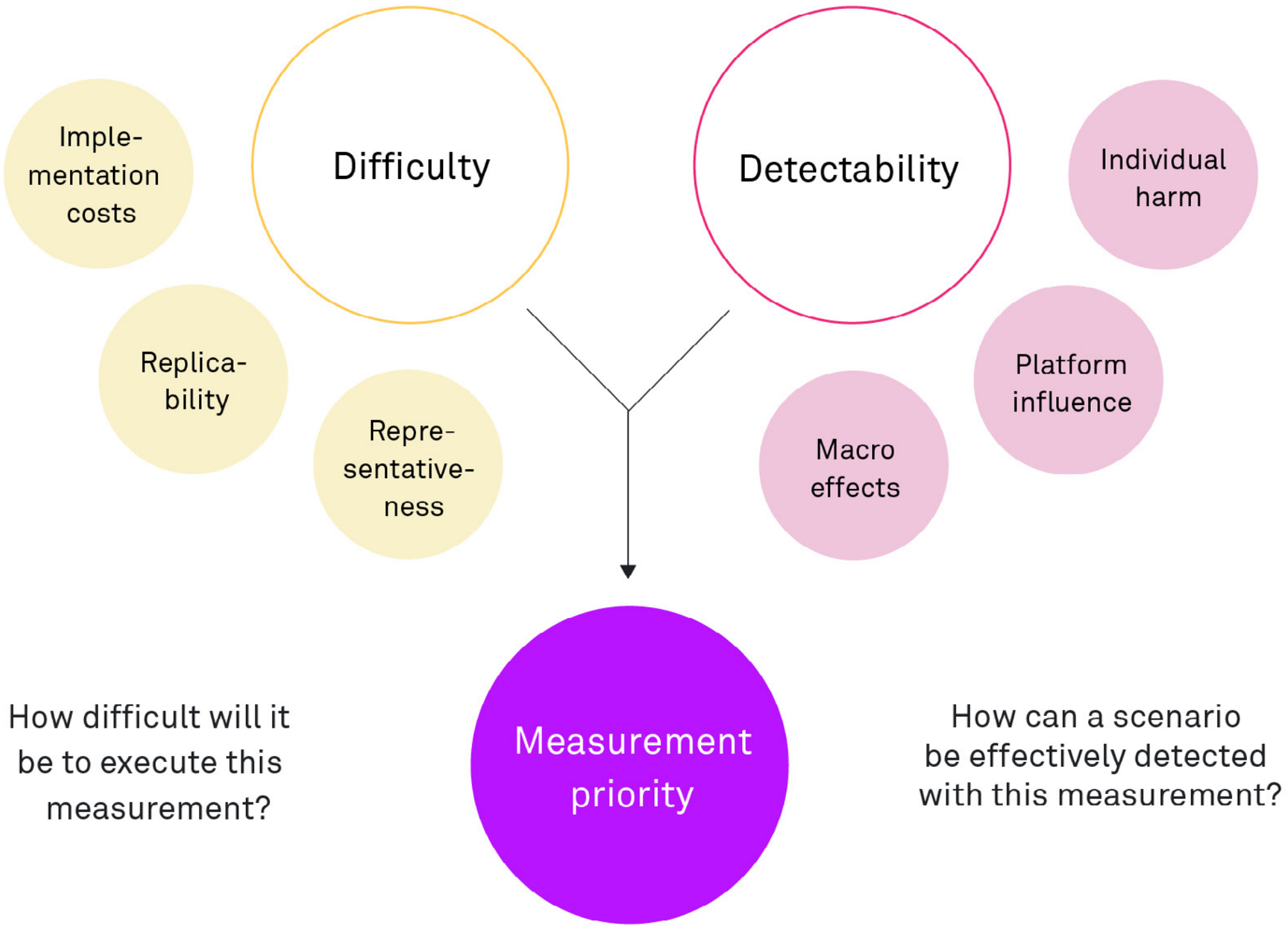

Figure 8:
Criteria to prioritise measurements.

In Section 7.2, we saw that, given the pervasiveness and complexity of risk scenarios, it was necessary to prioritise them to decide which one to investigate. The same is true for measurements. Studying a particular scenario may benefit from studying all the relevant platform elements involved, but investigating all of them is often not feasible. Once again, we need to prioritise; here, thinking of two variables may prove helpful: detectability and difficulty (see Figure 8).

**1. Detectability:** How can different scenario components be effectively detected with this measurement?

> Each measurement can usually only observe a part of the scenario: the individual harm, the specific recommender system's behaviour or the macro effects. Depending on the focus of the scenario, a measurement will help in understanding this impact.

**2. Difficulty:** How difficult will it be to execute this measurement?

> Especially concerning the implementation costs, replicability or representativeness, measurements will vary heavily from user studies requiring a careful study design and paying participants to document reviews based on simple online searches.

Ranking the ideas for measurements along these two parameters will provide an ordered list of the best ones that should be implemented to test a scenario. How many audit types are implemented and how many measurements are conducted depend on how good the chosen methods are and the available resources the audit team has.

We will examine the following risk scenario:

A young adult who is temporarily in a personal crisis is overly exposed to videos describing or showing self-harm by the recommender system on the 'For You' feed. This might exacerbate the general mental health crisis of young adults.

The platform element most interesting for this scenario is algorithmic logic. Regarding the types of audits, several approaches are possible to better understand the extent to which this scenario is happening on TikTok. After collecting potential approaches to test the scenario, you can rank them in terms of difficulty and detectability.

You could conduct a **data/scraping audit** to assess the extent of videos showing self-harm on TikTok. Therefore, you could inspect the platform's content either with first-party access to the data (e.g., getting metadata for all videos uploaded within a week) or by scraping content to get an idea of the number of distributed videos with self-harm content (e.g., scraping the search results for terms related to self-harm for 30 days). This would require a way to classify this type of content with some accuracy. What is considered self-harm content and what is not will likely demand human labelling. Accuracy can then be considered for assessing *detectability*, primarily concerning the influence of the recommender system and less so for the respective micro or macro harms. For example, the number of videos might only reflect that the macro effects exist but not whether the platform is contributing to it. The *difficulty* of this measurement depends on data access; it would be low if you have first-party access to existing classifications. Then, the measurements would also lead to representative and reproducible results. A scraping audit would be more difficult depending on the platform's restrictions. Analysis over time would increase the detectability of the macro effects by showing whether an increase in the number of videos precedes or follows an amplification of health crisis indicators. However, at the same time, a long-term analysis increases the difficulty.

A **sock puppet audit** could be conducted to check whether the recommender system produces excessive exposure, or so-called 'rabbit holes' (like the above-mentioned audit that The Wall Street Journal conducted). It could use accounts set up as adults between 18 and 25, scrolling the news feed for 60 minutes daily, pausing on videos (i.e., 'watching') related to depression and searching for related topics. The *detectability* hinges on two aspects: first, the accuracy of labelling this type of content, as described above, and, second, a definition for what level of exposure you consider 'excessive'. This type of audit will provide few insights into individual harm or macro effects, but the analysis could reveal how fast the recommender adapts to user behaviour, showing this type of content to users. Understanding how a specific product changes over time may also shed light on the capacity of the platform to react to overexposure and remedy it by implementing mitigation measures. The *difficulty* is likely higher than with the data/scraping audit because it takes more resources to implement automated behaviour comparable to real user interactions and scale them to a representative level. There is a systematic issue with the replicability of these audits. Over time, as the algorithm and content of the platform change, the same measurement could lead to different results.

With a **user and/or expert survey**, you could try to understand user experiences on the platform by surveying 1,000 regular TikTok users, shedding light on the potential individual harms they may suffer but also helping assess their macro effects. The *difficulty* of conducting representative surveys is generally high. Reaching the right audience for a survey might be particularly challenging for the scenario above because young adults could be more difficult to reach through existing surveying and sampling platforms. As with every other type of audit, it is essential to consider how representative its results are. Most findings will not be generalisable to what all the users of a platform experience. Therefore, auditors must be aware of the spectrum of applicability of their results and be fully transparent about them.

No single measurement will be perfect regarding all prioritisation criteria. Therefore, you need to find the right number of methods, likely more than one, that best fit the scenario and contribute to understanding the different components of the scenario. In this case, a data/scraping audit together with a user survey could already be sufficient to understand the impact of the platform and its recommender on individuals, as well as on macro harms.

## 8.4. Conducting measurements

Finally, the measurements you decided on must be implemented to collect the data. How exactly depends on the types of audits you prioritised and find most useful for the scenario you are evaluating. We recommend carefully reviewing the related work and examples of the existing measurements provided above before developing a measurement setup created from scratch. Doing so will save resources and increase replicability if you use or build upon existing tools and frameworks. For example, 4Cat[82] or TRex[83] are frameworks developed and maintained by other researchers and can serve as a starting point for your measurements. In any case, you should maintain proper documentation regarding the circumstances of each measurement, including the code and technical setup. This is necessary to ensure reproducibility or at least comparability with future measurements.

We recommend following research best practices when it comes to conducting your audit and archiving its results.[84] This will make it easier to analyse the data and report on the results in the final step of the audit process.

# 9. Step 4: Evaluating results and audit reporting

The final step of the audit process is to evaluate the results of the measurements, interpret them with respect to the scenario(s) and summarise everything in an audit report. For the evaluation, you need to compare your results with the expectations when selecting the type of audit. For example, did you reach the sample size you planned for in a survey? Did the automated audit result in a dataset allowing for statistical tests?

For the interpretation, you should reconsider the scenario and discuss to what extent your measurements confirm or contradict that the platform has an impact on the harms under scrutiny.

The documentation of the previous steps forms the basis of the report and allows the audit to be reproduced later, to allow replications by third-party auditors or to test if implemented interventions and mitigation measures work. Furthermore, revisiting scenarios and analysing them with other types of audits can also be useful for checking whether a mitigation measure has even addressed the different components of a scenario (e.g., individual harm and macro effect). Mitigations implemented in recommender systems might sometimes not reduce the harm itself but instead are optimised in such a

82 Digital Methods Initiative, '4CAT: Capture and Analysis Toolkit', Python, 9 January 2023, https://github.com/digitalmethodsinitiative/4cat.
83 Tracking Exposed, 'Tracking Exposed Toolkit', HTML, 8 January 2023, https://github.com/tracking-exposed/trex.
84 There are many guides available, and many organisations already have their own best practices. One very general guide can be found here: The Ohio State University, 'Research Data Management - Best Practices', accessed 15 December 2022, https://guides.osu.edu/rdm-best-practice/home.

way that the harmful content cannot be detected or measured anymore. For example, if you were studying representations of self-harm on a platform, the platform might simply ban specific hashtags as a mitigation measure. A replication study relying on measuring hashtags would then show that the prevalence of this content decreases, where instead users might have simply started using other hashtags, as is often the case with borderline or harmful content about suicide, drugs or political extremism.[85]

The DSA outlines only abstract requirements regarding the documentation of risk assessments but has some very specific requirements for compliance audit reports (see Infobox 'Documentation Requirements in the DSA regarding Audits and Risk Assessments'). Because it must be clarified by regulators if risk assessments will be part of the compliance audits and because the requirements for compliance audits are more concrete, we have incorporated the requirements for both in the following template. A report should provide detailed information to make the assessment reproducible and, if necessary, should also describe the mitigation measures the platforms should address. In the case of compliance audits, if the audit finds significant issues, VLOPs need to react to the recommendations addressed to them within a month (see Article 37(6)).

The content of an audit report depends on several factors. It matters whether it was carried out by a first, second or third party, what type of audit was performed and to which target audience it is addressed. However, we consider some elements fundamental for a thorough and transparent report. The following template summarises these elements:

**1. Executive Summary**

**a.** Describe your considerations of the prerequisite steps, including what platform is audited. Who is conducting the audit and why? What systemic risk(s) are considered for the audit?

**b.** A compliance audit needs to include names and addresses (as required for compliance audits, Article 37(4)(a)(b), see Infobox 'Documentation Requirements in the DSA regarding Audits and Risk Assessments').

**c.** Summarise the main findings (Article 37(4)(e)).

**d.** Compliance audits also need to include a declaration of interest (Article 37(4)(c)).

85 Taylor Lorenz, 'Internet "Algospeak" Is Changing Our Language in Real Time, from "Nip Nops" to "Le Dollar Bean"', *Washington Post*, 8 April 2022, https://www.washingtonpost.com/technology/2022/04/08/algospeak-tiktok-le-dollar-bean/.

**2. Introduction (Step 1)**

**a.** Describe details about the platform, for example, the products and elements relevant to the audit at the time. This is important because the platform might change between this audit and the next one. Elaborate on the following questions: What is the choice architecture of the platform, and what are its affordances? What elements of the platform use recommender systems? (Article 37(4)(d))

**b.** Describe the stakeholders considered for the audit and include a list of those involved in the process (Article 37(4)(f)).

**3. Scenarios (Step 2)**

**a.** Describe the process of scenario creation: How did you develop scenarios? What actors, characteristics, harms, platform involvement and macro impacts were considered? How was the list prioritised?

**b.** List the scenario(s) selected for the audit.

**4. Methods and Measurements (Step 3)**

**a.** Describe the process of measurement development: How did you develop measurements? What types of audits did you consider (e.g., with respect to the level of access you have to the platform)? What measurements were developed for the scenario(s)?

**b.** What methods were selected for implementation? Describe the reasoning behind your decision. Consider the difficulty and detectability of each measurement.

**5. Results and Audit Opinion (Step 4)**

**a.** Describe the findings of your audits with respect to the risk scenarios. Discuss to what extent your results can be generalised.

**b.** Provide a statement (i.e., 'audit opinion') on the systemic risks (Article 37(4)(g)).

**c.** Describe mitigations the platform should implement (Article 37(4)(h)).

**6. Appendix**

**a.** You should provide supplemental material necessary to understand or reproduce your audit. This can include surveys, additional results, documentation of the process and references to code or raw data.

**b.** When referencing external digital sources, ensure their long-term availability.

**Documentation requirements in the DSA regarding audits and risk assessments:**

**Recital (85)**
'In order to make it possible that subsequent risk assessments build on each other and show the evolution of the risks identified, as well as to facilitate investigations and enforcement actions, providers of very large online platforms and of very large online search engines should preserve all supporting documents relating to the risk assessments that they carried out, such as information regarding the preparation thereof, underlying data and data on the testing of their algorithmic systems.'

**Article 34(3)**
'Providers of very large online platforms and of very large online search engines shall preserve the supporting documents of the risk assessments for at least three years after the performance of risk assessments, and shall, upon request, communicate them to the Commission and to the Digital Services Coordinator of establishment.'

**Article 37(4)**
'Providers of very large online platforms and of very large online search engines shall ensure that the organisations that perform the audits establish an audit report for each audit. That report shall be substantiated, in writing, and shall include at least the following:
(a) the name, address and the point of contact of the provider of the very large online platform or of the very large online search engine subject to the audit and the period covered;
(b) the name and address of the organisation or organisations performing the audit;
(c) a declaration of interests;
(d) a description of the specific elements audited, and the methodology applied;
(e) a description and a summary of the main findings drawn from the audit;
(f) a list of the third parties consulted as part of the audit;
(g) an audit opinion on whether the provider of the very large online platform or of the very large online search engine subject to the audit complied with the obligations and with the commitments referred to in paragraph 1, namely "positive", "positive with comments" or "negative";
(h) where the audit opinion is not "positive", operational recommendations on specific measures to achieve compliance and the recommended timeframe to achieve compliance.'

# 10. Summary and outlook

As described in the beginning, we face two challenges in auditing and assessing recommender systems and their risks. First, social media platforms and their recommender systems are composed of multiple products that are embedded in a specific choice architecture and shaped by platform-specific affordances. Second, systemic risks, as mentioned in the DSA, are often described vaguely and abstractly.

The DSA now requires various procedures to audit algorithmic systems and assess how they contribute to systemic risks. In our paper, we have identified different audits and assessments as demanded by the DSA (see Section 4.1):

- First-party 'risk assessments' are conducted by VLOPs internally (i.e., Article 34, DSA) and published once a year.
- 'DSA compliance audits' are a type of second-party audit conducted by external contractors. It is unclear whether these include any technical aspects.
- Third-party 'risk assessments' or 'impact assessments' are carried out by external and independent researchers (i.e., Article 40, DSA).

These required audits and risk assessments are—albeit important—snapshots because the platforms and the way users engage with content continuously evolve. Furthermore, to investigate systemic risks, you must bear in mind that they are an interplay between individual and societal phenomena, with platform-specific risks induced by their affordances and recommender systems. This tension and the intrinsic complexity of social and psychological phenomena require that various relevant parties participate in the audit. For this reason, we propose a multistakeholder audit process breaking down the abstract systemic risks into concrete scenarios and then conducting multiple types of audits to ensure that the effects within and outside the platform are studied (e.g., by combining user surveys or expert interviews with an automated audit).

The present paper provides guidelines to operationalise systemic risks, set up a process to make well-founded decisions on suitable audit methods and enable comparability between different approaches. The risks-scenario-based audit process meets requirements for first-, second- and third-party audits. It helps auditors to set up an audit process which brings together various elements, issues, stakeholders and audit types. The audit process is based on four steps described in Sections 5–9 and explains how the different components are related to DSA requirements. Our guidelines can also help you revisit scenarios after a change to the platform has been implemented, allowing you to assess whether the risk mitigations deployed have had the desired effect.

We explicitly suggest that the risk-scenario-based audit process we have put forward in this paper be considered a contribution to the ongoing discussions on how the DSA should be implemented. At the time of writing this, several pieces of secondary legislation were being developed. Among them were an implementing act on a transparency reports template, a delegated act on independent external audits and guidelines on specific risks related to risk mitigation.[86]

Below are some recommendations we offer for the enforcement of the DSA. We detected the issues these recommendations seek to address while developing the risk-scenario-based approach for audits of social media platforms and during our exchanges with experts and other stakeholders.

**Prevent 'audit-washing' through compliance audits**

The DSA requires independent contractors to audit if VLOPs violate their due diligence obligations. Therefore, the European Commission is currently working on the delegated act on independent audits according to Article 37—the audits we call compliance audits. An internal presentation by the European Commission suggests that these independent compliance audits should be part of a life cycle of risk management supervision by the European Commission. Another part of this life cycle is risk assessments, which, in turn, are also an object of scrutiny for compliance audits. However, it is still unclear who these independent auditors are going to be and if compliance audits must encompass the auditing of the VLOPs' internal risk assessments for quality.

Therefore, we recommend the following three measures:

1. Make sure that independent compliance audits of external contractors also evaluate the quality and methods of the VLOPs' internal risk assessments.
2. Ensure that these independent second-party audits include independent researchers and experts on systemic risks and algorithm audits as stakeholders in their auditing processes.
3. Ideally, second-party audits include independent risk assessments themselves using the risk scenario audit process that we have described as a facilitating tool.

These requirements are essential to ensure that second-party audits are not rendered mere 'audit-washing' tools for the platforms, as the audit

86 eCommerce Experts Group, '25th ECommerce Expert Group Meeting'.

researchers Ellen Goodman and Julia Trehu suggested.[87]

**Extended transparency and observability of risk assessments are necessary**

The topic of data access and transparency is worthy of a paper of its own, which is why we cannot go in-depth on this issue here. However, we briefly want to discuss several points with strong connections to risk assessments and audits in the following sections.

As we outlined in Section 4.1, VLOPs are obliged under Article 42(4)(a) to report the results of their internal risk assessments to the digital services coordinator of establishment and the European Commission in accordance with Article 34 and to make them publicly available.

For transparency and to enable their scrutiny by any interested party, VLOPs should also be required to provide a thorough methodological description of the assessments they have carried out, including their hypotheses, information about which metrics were considered and if and how they assessed their own internal experiments (such as A/B-testing). Making this available to the public would enable reviews of these first-party audits by third-party auditors, which can only result in heightened accountability. This is already vaguely stated in Article 42. However, we advocate for these audit processes, especially the tested scenarios, to be disclosed.

**Observability should be the goal of transparency and data access obligations**

As described above, platforms and how users engage with them continuously evolve. Therefore, the information required to conduct audits and, thus, understand the impact that regulations and their enforcement have had on risk mitigation measures is also constantly changing. If transparency and data access obligations do not reflect the dynamic nature of the issue and instead focus on static metrics, platforms might try to optimise these numbers instead of actually mitigating risks.

Therefore, the goal of transparency and data access obligations should be to maximise observability, that is, our capacity to scrutinise these platforms over time, their processes and, thus, being able to detect their evolution and impacts on individuals, groups and society—also across different platforms.

**This is why, data access needs to include real-time data**. Although this is already mentioned in the DSA (e.g., Recital 98), platforms can backtrack if providing access is technically complicated. Instead, platforms should be

87 Ellen P. Goodman and Julia Trehu, 'AI Audit Washing and Accountability', SSRN Scholarly Paper (Rochester, NY, 22 September 2022), https://doi.org/10.2139/ssrn.4227350.

required to give access to and provide the technical means for auditors to analyse the platforms in real time. This must be further specified in the delegated and implementing acts because this is an important aspect towards achieving platform observability.

This also shows that **transparency reports need standards**. Experiences with the transparency reports under the Network Enforcement Act in Germany have shown that these reports are difficult to compare and assess, for example, because they do not disclose the societal scale[88] or explain *why* these harmful exposures are happening. Also, these reports sometimes work with different metrics, even within one report, or do not disclose the basic population of the provided prevalence numbers. We recommend more precise specifications for DSA transparency reports (e.g., in accordance with the European Digital Media Observatory (EDMO)[89] and Algorithm Watch[90]). In line with this, we strongly support the recommendation by the Integrity Institute, an American nonprofit organisation working to improve social platforms, calling for these reports to disclose the entire life cycle of harmful content.[91]

### Maintain critical supervision

The European Commission and Digital Service Coordinators are responsible for the supervision of transparency and audit reports. We worry that any oversight approach based on key performance indicators or similar metrics (e.g., focusing on the number of deleted posts or comparable high-level data in dashboards) might distract from what should be the focus of interest: actual risks and specific scenarios. Therefore, we call for quality-based approaches to supervision, which not only check whether risk assessments were conducted, but also evaluate them.[92]

As a civil society organisation, we want to contribute to setting high standards for the societal scrutiny of the recommender systems of social networks and

88 Integrity Institute, 'Metrics & Transparency. Data and Datasets to Track Harms, Design, and Process on Social Media Platforms (Draft)', 22 September 2021, https://static1.squarespace.com/static/614cbb3258c5c87026497577/t/617834d31bcf2c5ac4c07494/1635267795944/Metrics+and+Transparency+-+Summary+%28EXTERNAL%29.pdf.

89 Marta Cantero Gamito et al., 'Report of the European Digital Media Observatory's Working Group on Platform-to-Researcher Data Access' (Brussels: European Digital Media Observatory, 31 May 2022), https://edmo.eu/wp-content/uploads/2022/02/Report-of-the-European-Digital-Media-Observatorys-Working-Group-on-Platform-to-Researcher-Data-Access-2022.pdf.

90 John Albert, 'A Guide to the EU's New Rules for Researcher Access to Platform Data', *AlgorithmWatch*, 7 December 2022, https://algorithmwatch.org/en/dsa-data-access-explained/; John Albert et al., 'Policy Brief: Our Recommendations for Strengthening Data Access for Public Interest Research' (AlgorithmWatch, 5 July 2022), https://algorithmwatch.org/en/policy-brief-platforms-data-access/.

91 Integrity Institute, 'Metrics & Transparency. Data and Datasets to Track Harms, Design, and Process on Social Media Platforms (Draft)'.

92 Julian Jaursch, 'Platform Oversight: Here Is What a Strong Digital Services Coordinator Should Look Like', *Verfassungsblog*, 31 October 2022, https://verfassungsblog.de/dsa-dsc/.

similar online platforms. Audits that are mainly conducted internally as first-party audits and only reviewed for compliance by second-party auditors will fail to effectively minimise the systemic risks that these technologies pose. Therefore, we see our approach as a first attempt at standardising the audit processes already present in the DSA, one that seeks to achieve comparability. We believe that the approach outlined in this document can contribute to furthering the operationalisation of the audits and risk assessments that the DSA requires and those carried out beyond its scope. Moreover, we hope it can also become part of the toolbox that every concerned organisation or authority has at hand to watch over the pressing challenges of our constantly evolving digital public sphere.

## Acknowledgements

We would like to thank the experts from civil society, academia, companies and authorities with whom we developed ideas and who gave feedback on our paper. We also wish to thank our workshop participants for testing our risk-scenario-based audit process. We greatly appreciated criticism and comments from John Albert (AlgorithmWatch), Sara Bundtzen (Institute for Strategic Dialogue), Claudio Agosti (Tracking Exposed), Marcus Bösch (TikTok Newsletter), Jenny Brennan (Ada Lovelace Institute), Ciarán O'Connor (Institute for Strategic Dialogue), Maximilian Gahntz (Mozilla), Brandi Geurkink (Mozilla; Reset), Bernhard Rieder (University of Amsterdam), Salvatore Romano (Tracking Exposed) and Matthias Spielkamp (AlgorithmWatch).

The Stiftung Neue Verantwortung (SNV) team deserves great recognition for their excellent support during the entire research and publication process: Alexander Hohlfeld, Santiago Sordo Ruz, Julian Jaursch, Luisa Seeling, Corbinian Ruckerbauer, Anna Semenova, Alina Siebert and Stefan Heumann.

The views expressed in the text do not necessarily reflect those of the experts with whom we have exchanged views or their employers, and any remaining errors are our own.

The paper is part of a project entitled 'Approaches to Analyse and Evaluate AI-Based Recommendation Systems for Internet Intermediaries', which is funded by the German Beauftragte der Bundesregierung für Kultur und Medien (BKM).

## About the Stiftung Neue Verantwortung

The Stiftung Neue Verantwortung (SNV) is an independent, nonprofit think tank working at the intersection of technology and society. SNV's core method is collaborative policy development, involving experts from government, tech companies, civil society and academia to test and develop analyses with the aim of generating ideas on how governments can positively shape the technological transformation. To guarantee the independence of its work, the organisation has adopted a concept of mixed funding sources that include foundations, public funds and corporate donations.

# About the authors

**Dr Anna-Katharina Meßmer** is the SNV's project director for 'Strengthening the Digital Public Sphere and Platform Regulation' and works on approaches to analyse and evaluate AI-based recommendation systems for internet intermediaries. Her research is based on technical, sociological and policy perspectives.

From 2019–2021, she led the widely acclaimed study on digital news and information literacy 'Source: the Internet'. Previously, she was managing director of the 'Research Institute for Societal Development' and worked as Lead Strategic Development at the opinion research start-up Civey. As a researcher, Meßmer was responsible for projects on political and social issues at the Institute for Media and Communication Policy at the Technical University of Berlin and the Ludwig-Maximilians-University of Munich.

**Dr Martin Degeling** joined SNV in May 2022 to work on the project to analyse and evaluate AI-based recommendation systems for internet intermediaries.

He is interested in the (semi) automated analysis and auditing of IT systems regarding their effects on individuals and societies. Until June 2022, he was scientific coordinator for a research training group on human-centred systems security at Ruhr University Bochum and was previously a postdoctoral fellow at Carnegie Mellon University in Pittsburgh. In his research, he analysed how legal privacy frameworks were adopted on websites, showing which technical and organisational challenges occurred. He has also examined how behavioural data are collected and leveraged to create profiles on users.

**Contact the Authors:**

Anna-Katharina Meßmer
kmessmer@stiftung-nv.de

Martin Degeling
mdegeling@stiftung-nv.de

# Imprint